\documentclass[acmsmall,screen]{acmart}

\AtBeginDocument{%
  }

\setcopyright{rightsretained}
\acmJournal{PACMHCI} 
\acmYear{2024} 
\acmVolume{8} 
\acmNumber{CSCW1} 
\acmArticle{206} 
\acmMonth{4}
\acmDOI{10.1145/3653697}

\usepackage{graphics}      
\usepackage{xcolor}
\usepackage{booktabs}
\usepackage{multirow}
\usepackage{enumitem}
\usepackage{upquote}
\usepackage{csquotes}
\renewcommand{\mkbegdispquote}[2]{\itshape}
\AtBeginEnvironment{quote}{\itshape}
\usepackage{tikz}
\usetikzlibrary{decorations.text,calc,arrows.meta}
\usetikzlibrary {arrows.meta,bending,positioning,fit}
\usepackage{mdframed}
\usepackage{smartdiagram}
\usepackage[font=small]{caption}
\usepackage{subcaption}
\usepackage{scalerel,graphicx,xparse}
\colorlet{BLUE}{blue}

\usepackage{natbib}

\usepackage{hyperref}
\usepackage[capitalise, nameinlink]{cleveref}

\usepackage{todonotes}
\usepackage{listings}
\usepackage{rotating}

\definecolor{seabornBlue}{RGB}{76,114,176}
\definecolor{seabornGreen}{RGB}{85,168,104}
\definecolor{seabornRed}{RGB}{196,78,82}
\definecolor{seabornOrange}{RGB}{221, 132, 82}

\definecolor{infra}{RGB}{1, 115, 178}
\definecolor{component}{RGB}{222, 143, 5}
\definecolor{pipeline}{RGB}{2, 158, 115}
\definecolor{run}{RGB}{213, 94, 0}

\definecolor{datacoll}{RGB}{204, 120, 188}
\definecolor{expt}{RGB}{202, 145, 97}
\definecolor{evaldeploy}{RGB}{251, 175, 228}
\definecolor{monitorresp}{RGB}{148, 148, 148}

\colorlet{punct}{red!60!black}
\definecolor{background}{HTML}{EEEEEE}
\definecolor{delim}{RGB}{20,105,176}
\colorlet{numb}{magenta!60!black}

\newcommand{\topic}[1]{\vspace{-3.5pt}\smallskip \smallskip \noindent{\bf #1.}}

\usepackage[normalem]{ulem}
\usepackage{cancel}
\newcommand\redline{\bgroup\markoverwith
    {\textcolor{red}{\rule[.5ex]{2pt}{0.4pt}}}\ULon}

\colorlet{revisionBlue}{black}

\newcommand{\addition}[1]{{\color{revisionBlue}{#1}}}
\newcommand{\additiontwo}[1]{{\color{black}{#1}}}

\makeatletter
\def\url@leostyle{%
  \@ifundefined{selectfont}{
    \def\UrlFont{\sf}
  }{
    \def\UrlFont{\small\bf\ttfamily}
  }}
\makeatother
\definecolor{linkColor}{RGB}{6,125,233}
\definecolor{linkpurple}{HTML}{902043}

\begin{document}

\title[{\em``We have no idea how models will behave in production until production''}: How engineers operationalize ML]{{\em``We Have No Idea How Models will Behave in Production until Production''}: How Engineers Operationalize Machine Learning}

\author{Shreya Shankar}
\authornote{Both authors contributed equally to this research.}
\affiliation{%
  \institution{University of California, Berkeley}
  \city{Berkeley}
  \state{CA}
  \country{USA}
}
\email{shreyashankar@berkeley.edu}

\author{Rolando Garcia}
\authornotemark[1]
\affiliation{%
  \institution{University of California, Berkeley}
  \city{Berkeley}
  \state{CA}
  \country{USA}
}
\email{rogarcia@berkeley.edu}

\author{Joseph M. Hellerstein}
\affiliation{%
  \institution{University of California, Berkeley}
  \city{Berkeley}
  \state{CA}
  \country{USA}
}
\email{hellerstein@berkeley.edu}

\author{Aditya G. Parameswaran}
\affiliation{%
  \institution{University of California, Berkeley}
  \city{Berkeley}
  \state{CA}
  \country{USA}
}
\email{adityagp@berkeley.edu}

\renewcommand{\shortauthors}{Shankar \& Garcia et al.}

\begin{abstract}
  Organizations rely on machine learning engineers (MLEs) to deploy models and maintain ML pipelines in production.
  Due to models' extensive reliance on fresh data, the operationalization of machine learning, or MLOps,
  requires MLEs to have proficiency in data science and engineering.
  When considered holistically, the job seems staggering---how do MLEs do MLOps,
  and what are their unaddressed challenges?
  To address these questions, we conducted semi-structured ethnographic interviews with 18 MLEs
  working on various applications, including chatbots, autonomous vehicles, and finance.
  We find that MLEs engage in a workflow of (i) data preparation, (ii) experimentation,
  (iii) evaluation throughout a multi-staged deployment, and (iv) continual monitoring and response.
  Throughout this workflow, MLEs collaborate extensively with data scientists,
  product stakeholders, and one another, supplementing routine verbal exchanges
  with communication tools ranging from Slack
  to organization-wide ticketing and reporting systems.
  We introduce the 3Vs of MLOps: \textit{velocity}, \textit{visibility}, and \textit{versioning}---three
  virtues of successful ML deployments that MLEs learn to balance and grow as they mature.
  Finally, we discuss design implications and opportunities for future work.
\end{abstract}



\received{15 January 2023}


\maketitle

\section{Introduction}
\label{sec:intro}

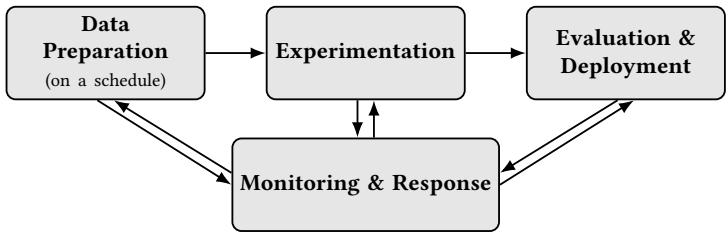
\begin{figure}
    \centering
    \begin{center}
    \resizebox{0.7\linewidth}{!}
        {\tikzstyle{n} = [rectangle, draw, fill=gray!20, node distance=4cm, text width=8em, text centered, rounded corners, minimum height=4em, thick]
        \tikzstyle{l} = [draw, -Latex, thick]
        \begin{tikzpicture}
            \begin{scope}[local bounding box=misc]
                \node[n] (A) {{\bf Data}\\{\bf Preparation} {\footnotesize (on a schedule)}};
                \node[n, right of=A] (B) {{\bf Experimentation}};
                \node[n, right of=B] (C) {{\bf Evaluation \& Deployment}};
            \end{scope}
            \node[n, below of=B, node distance=2cm, text width=11em] (D) {{\bf Monitoring \& Response}};

            \path [l] (A) -- (B);
            \path [l] (B) -- (C);

            \path [l] (A.260) -- (D.180);
            \path [l] (D.175) -- (A.280);

            \path [l] (B.260) -- (D.100);
            \path [l] (D.80) -- (B.280);

            \path [l] (C.260) -- (D.5);
            \path [l] (D.0) -- (C.280);


        \end{tikzpicture}}
    \end{center}
    \caption{Core tasks in the MLOps workflow. Prior work discusses a production data science workflow of preparation, modeling, and deployment~\cite{autoai}. Our work exposes (i) the scheduled and recurring nature of \textbf{data preparation} (including automated ML tasks, such as model retraining), identifies (ii) a  broader \textbf{experimentation} step (which could include modeling or adding new features), and provides more insight into human-centered (iii) \textbf{evaluation \& deployment}, and (iv) \textbf{monitoring \& response}.}
    \label{fig:tasksdiagram}
\end{figure}

As machine learning (ML) models are increasingly
incorporated into software,
a nascent sub-field called {\em MLOps} (short for ML Operations) has emerged
to organize the
``set of practices that aim to deploy and maintain ML
models in production reliably and efficiently''~\cite{enwiki:1109828739,alla2021mlops}.
It is widely recognized that \addition{MLOps issues pose challenges to organizations}.
Anecdotal reports claim that 90\% of ML models
don't make it to production~\cite{weiner2020ai};
others claim that 85\% of ML projects fail to deliver value~\cite{ai-investments}\addition{---signaling the fact that translating ML models to production is difficult}.

At the same time, it is unclear {\em why} MLOps \addition{issues are difficult to deal with}.
Our present-day understanding of MLOps is limited to a fragmented landscape of white papers, anecdotes, and
thought pieces~\cite{eric, mlreef_2021, garcia2018context, ghosh_2021},
as well as a cottage industry of startups aiming to address MLOps issues~\cite{huyen_2020}.
\addition{Early work by~\citet{Sculley2015HiddenTD}
attributes MLOps challenges to {\em technical debt}, analogous to that in software engineering but exacerbated in ML.}
Prior work has studied general practices of
data scientists working on \addition{ML}~\cite{sambasivan2021everyone, kandel, muller2019data, zhang2020data},
but successful ML deployments seem to further involve
a ``team of engineers who spend a
significant portion of their time on the less
glamorous aspects of ML like maintaining and monitoring
ML pipelines''
---that is, ML engineers (MLEs)~\cite{polyzotis2017data}.
It is well-known that MLEs typically need to have strong data science
and engineering skills~\cite{amershi_software_2019},
but it is unclear how those skills are used in their day-to-day workflows.

There is thus a pressing need to bring clarity to MLOps---specifically
in identifying what MLOps typically involves---across organizations and ML applications.
\addition{While papers on MLOps have described specific case studies, prescribed best practices, and surveyed tools to help automate the ML lifecycle, there is a pressing need to understand the {\em human-centered} workflow required to support and sustain the deployment of ML models in practice. A richer understanding of \addition{common} practices and challenges in MLOps can surface gaps
in present-day processes and better inform the development of next-generation ML engineering tools.}
\addition{To address this need,} we conducted a semi-structured interview study of \addition{ML engineers (MLEs),}
each of whom has \addition{been responsible for a production ML model.}
\addition{With the intent of identifying common themes across organizations and industries, we sourced 18 ML engineers from different companies and applications,}
and asked them open-ended questions to understand their workflow and day-to-day
challenges---both on an individual and organizational level.

Prior work \addition{focusing on the earlier stages of data science has shown that it is a largely iterative and manual process, requiring humans to perform several stages
of data cleaning}, exploration, model building,
and visualization~\cite{passi2018trust, krossorienting,
    sambasivan2021everyone, hohman2020understanding}.
Before embarking on our study,
we expected that \addition{the subsequent deployment of ML models}
in production would instead be more amenable to automation,
with less need for human intervention and supervision.
Our interviews, in fact, revealed the opposite---much like
the earlier stages of data science,
deploying and maintaining models in production
is highly iterative, manually-intensive, and team-oriented.
Our interviewees emphasized organizational and collaborative strategies
to sustain ML pipeline performance and minimize
pipeline downtime, mentioning on-call rotations,
manual rules and guardrails,
and teams of practitioners inspecting data quality alerts.

In this paper, we provide insight into human-centered aspects of MLOps practices
and identify opportunities for future MLOps tools. \addition{We conduct a semi-structured interview study solely focused on ML engineers, an increasingly important persona in the broader software development ecosystem as more applications leverage ML. Our focus on MLEs, and uncovering their workflows and challenges as part of the MLOps process, addresses a gap in the literature.} Through our interviews, we characterize
an ML engineer's typical workflow \addition{(on top of automated processes)} into four stages (\Cref{fig:tasksdiagram}):
(i) data preparation,
(ii) experimentation,
(iii) evaluation and deployment, and
(iv) monitoring and response,
all centered around team-based, collaborative practices.
Key takeaways for each stage are as follows:

\topic{Data ingestion often runs automatically,
    but MLEs drive data preparation through data selection,
    analysis, labeling, and validation (\Cref{sec:summary-dataprep})}
We find that organizations typically leverage teams of data engineers
to manage recurring end-to-end executions of data pipelines, allowing MLEs
to focus on ML-specific steps such as defining features,
a retraining cadence, and a labeling cadence.
If a problem can be automated away, engineers prefer to do
so---e.g., retraining models on a regular cadence to protect against changes
in the distribution of features over time.
Thus, they can spend energy on tasks that require human input,
such as supervising crowd workers who provide input labels or resolve inconcistencies in these labels.

\topic{Even in production, experimentation is highly iterative and collaborative,
    despite the use of model training tools and infrastructure (\Cref{sec:summary-expt})}
As mentioned earlier, various articles claim that it is a problem for 90\% of models to never make
it to production~\cite{weiner2020ai},
but we find that this statistic is misguided. The nature of constant experimentation is bound to create many versions of models,
a small fraction of which (i.e. ``the best of the best'')
will make it to production.
MLEs discussed exercising judgment when choosing next experiments to run,
and expressed reservations about AutoML tools,
or ``keeping GPUs warm,'' given the vast search space.
MLEs consult domain experts and stakeholders in group meetings,
and prefer to iterate on the data (e.g., to identify new feature ideas) over innovating on model architectures.

\topic{Organizations employ a
    multi-stage model evaluation and deployment process,
    so MLEs manually review and authorize deployment to subsequent stages (\Cref{sec:summary-eval})}
Textbook model evaluation ``best practices'' do not do justice to the rigor with which organizations
think about deployments: they generally focus on using one typically-static held-out dataset in an offline manner to evaluate the model on~\cite{lin2021the}
and a single ML metric choice (e.g., precision, recall)~\cite{courseraeval}.
We find that many MLEs carefully deploy changes to increasing fractions of the population in stages.
At each stage, MLEs seek feedback from other stakeholders (e.g., product managers and domain experts) and invest significant resources in maintaining multiple up-to-date evaluation
datasets and metrics over time---especially to ensure that data sub-populations of interest are adequately covered.

\topic{MLEs closely monitor deployed models and stand by,
    ready to respond to failures in production (\Cref{sec:summary-monitoring})}
MLEs ensured that deployments were reliable via strategies such as on-call rotations, data monitoring,
and elaborate rule-based guardrails to avoid incorrect outputs.
MLEs discussed pain points such as alert fatigue from alerting systems and
the headache of managing pipeline jungles~\cite{Sculley2015HiddenTD},
or amalgamations of various filters, checks, and data transformations added to ML pipelines over time.

\addition{The rest of our paper is organized as follows: we cover background and work related to MLOps from the CSCW, HCI, ML, software engineering, and data science communities (\Cref{sec:related}). Next, we describe the methods used in our interview study (\Cref{sec:methods}). Then, we present our results and discuss our findings, including opportunities for new tooling (\Cref{sec:summary} and \Cref{sec:discussion}). Finally, we conclude with possible areas for future work.}

\section{Related Work}
\label{sec:related}

Our work builds on previous studies of data
and ML practitioners in industry.
We begin with the goal of characterizing the role of
an ML Engineer, starting with related data science
roles in the literature and drawing distinctions that
make MLEs unique. \addition{We then review work that discusses data science and ML workflows, not specific to MLOps. Third, we cover challenges
that arise from productionizing ML. Fourth, we survey
software engineering practices in the literature that tackle such challenges. Finally, we review recent work that explicitly attempts to define and discuss MLOps practices.}

\subsection{Characterizing the ML Engineer}\label{sec:related-dsroles}

Data science roles span various engineering and research tasks~\cite{kim2016}, and
many data-related activities are performed by people without
``data'' or ``ML'' in their job title~\cite{kim2017data},
so it can be hard to clearly define job descriptions~\cite{muller2019data}.
Nonetheless, since we focus on production ML pipelines, we discuss personas related
to data science, ML, and engineering---culminating in the description of the persona we study.

\vspace{1em}\noindent{\textbf{The Data Scientist:}}
Multiple studies have identified subtypes of data scientists,
some of whom are more engineering-focused than
others~\cite{zhang2020data, kim2016}. \citet{zhang2020data}
describe the many roles data scientists can take---communicator,
manager/executive, domain expert, researcher/scientist, and engineer/analyst/programmer.
They found considerable overlap in skills and tasks performed between
the (i) engineer/analyst/programmer and (ii) researcher/scientist roles:
both are highly technical and collaborate extensively.
Separately, Kim et al. taxonomized data scientists as:
insight providers, modeling specialists, platform builders,
polymaths, and team leaders~\cite{kim2016}. Modeling specialists
build predictive models using ML, and platform builders balance both
engineering and science as they produce reusable software across products.

\vspace{1em}\noindent{\textbf{The Data Engineer:}}
While data scientists engage in activities like exploratory
data analysis (EDA), data wrangling, and insight
generation~\cite{kandel, Wongsuphasawat2019GoalsPA}, \textit{data engineers}
are responsible for building robust pipelines that regularly transform and
prepare data~\cite{udacity_2020}. Data engineers often have a software
engineering and data systems background~\cite{kim2016}. In contrast, data scientists
typically have modeling, storytelling, and mathematics backgrounds~\cite{kim2016, udacity_2020}.
Since production ML systems involve data pipelines and ML
models in larger software services, they require a combination of
data engineering, data science, and software engineering skills.

\vspace{1em}\noindent{\textbf{The ML Engineer (MLE):}}
Our interview study focuses on the distinct \emph{ML Engineer}
(MLE) persona. MLEs have a multifaceted skill set:
they know how to transform data as inputs to ML pipelines,
train ML models, serve models, and wrap these pipelines in software
repositories~\cite{mlopsoda, john2021towards, renggli2021data}.
MLEs need to regularly process data at scale (much like data engineers~\cite{mlopsoda}),
employing statistics and ML techniques as do data scientists~\cite{Patel2008InvestigatingSM},
and are responsible for production artifacts as are software engineers~\cite{leite2019survey}.
Unlike typical data scientists and data engineers,
MLEs are responsible for deploying ML models and
maintaining them in production.

We classify production ML into two modes.
One, which we call \emph{single-use} ML, is more client-oriented,
where the focus is to generate predictions once to make a
specific data-informed business decision~\cite{krossorienting}.
Typically, this involves producing reports, primarily performed by data
scientists~\cite{hackingwithnpos, kim2017data}. In the other mode, which
we call \emph{repeated-use} ML, predictions are repeatedly generated, often
as intermediate steps in data pipelines or as part of ML-powered products, such
as voice assistants and recommender systems~\cite{kim2016,amershi_software_2019}.
Continuously generating ML predictions over time requires more data and software engineering
expertise~\cite{kim2017data, tamburri2020sustainable}. In our study, we focus on MLEs
who work on the latter mode of production ML.

\subsection{Machine Learning Workflows}\label{sec:related-mlworkflows}

Here, we cover literature on ML practitioners'
workflows. We discuss both technical and collaborative workflows,
and then we describe the workflow our study seeks to uncover.


Several studies have investigated aspects of the broader ML workflow, mostly in single-use production ML applications. \addition{Early studies on data science workflows point to industry-originated software development process models, such as the Agile framework~\cite{cohen2004introduction} and the CRoss Industry Standard Process for Data Mining (CRISP-DM)~\cite{chapman1999crisp}. More recently,~\citet{studer2021towards} introduce CRISP-ML, a new process model that augments CRISP-DM with a final ``monitoring and maintenance'' phase to support ML workflows.} \citet{muller2019data} interview practitioners and focus on the data practices of
data science workflows, breaking them down into discovery, capture, design, and curation.
\citet{Wongsuphasawat2019GoalsPA}'s workflow includes some ML: it consists of data acquisition,
wrangling, exploration, modeling, and reporting. \citet{autoai} takes another step back
and includes productionization; they identify three high-level phases of preparation, modeling,
and deployment. Preparation includes activities ranging from wrangling~\cite{kandel2011wrangler} to
feature engineering~\cite{Patel2008InvestigatingSM}. Modeling includes selection, hyperparameter optimization,
ensembling, and validation, and deployment includes monitoring and improvement~\cite{autoai,zhang2020data}.
Of the three large stages, several studies have identified preparation as the most time-intensive
stage of the workflow~\cite{muller2019data,guowrangler}, where data scientists commonly iterate on rules to help
generate features~\cite{Passi2017DataVL, hohman2020understanding}.

While the above stages of the data science workflow comprise a loop of technical tasks, \citet{krossorienting}
identify an \emph{outer loop} of data science, centered around collaborative practices.
The outer loop consists of groundwork (i.e., building trust), orienting,
problem framing, magic (i.e., technical loop), and counseling. While \citet{krossorienting}'s
loop focuses on data science work that directly interacts with clients, mostly in the form of single-use
ML, similar themes emerge when performing repeated-use ML engineering work, e.g., repeatedly generating ML
predictions in an automated fashion. In production settings, predictions must yield value for the
business~\cite{Patel2008InvestigatingSM}, requiring some groundwork, orienting, and problem framing.
In their paper on tensions around collaborative, applied data science work, \citet{passi2018trust}
discuss that it's important to align different stakeholders on system performance metrics:
for example, one of their interviewees mentioned that accuracy is a ``problematic'' metric because
different users interpret it differently. In another example, \citet{holstein_improving_2019} say
that a single global metric doesn't capture performance for certain groups of users
(e.g., accuracy for a subgroup might decrease when overall accuracy increases).

In our study, we characterize the workflow from a repeated-use ML engineering perspective, focusing
on specific practices within deployment stages. \addition{Some related work defines steps in the ML workflow, such as model training and model monitoring, through both short papers~\cite{matsui2022mlops} and extensive literature reviews~\cite{kumara2022requirements}. We take a different but complementary approach: like \citet{muller2019data} who focus on data scientists, we conduct an interview study of MLEs, using grounded theory to analyze our findings~\cite{strauss1994grounded}.}
Further, our study seeks to uncover collaborative practices and challenges, focusing on the ML engineering perspective,
and how MLEs align all stakeholders such that ML systems continually generate value.

\subsection{Production ML Challenges}\label{sec:related-prodmlchallenges}

\citet{Sculley2015HiddenTD} were early proponents that production ML systems raise special
challenges and can be hard to maintain over time, based on their experience at Google.
They coined the ``Changing Anything Changes Everything'' (CACE) principle:
if one makes a seemingly innocuous change to an ML system,
such as changing the default value for a feature from 0 to -1,
the entire system's behavior can drastically change.
CACE easily creates technical debt and is often exacerbated
as errors ``cascade,'' or compound, throughout an end-to-end
pipeline~\cite{Sculley2015HiddenTD, sambasivan2021everyone,deploychallenges,polyzotis_data_2018,polyzotis2017data}.
We cover three well-studied challenges in production ML: data quality, reproducibility, and specificity.

\addition{First, ML predictions are only as good as their input data~\cite{polyzotis_data_2018, polyzotis2017data},
requiring active efforts from practitioners to ensure good data quality~\cite{dataval}}.
\citet{xin2021production} observe that production ML pipelines consist of models that are automatically retrained,
and we find that this retraining procedure is a pain point for practitioners because it requires constant monitoring of data.
If a model is retrained on bad data, all future predictions will
be unreliable. Data distribution shift is another known data-related
challenge for production ML systems~\cite{failingloudly, Ovadia2019CanYT, unifyingview, Wiles2021AFA, sugiyama},
and our work builds on top of the literature by reporting on how practitioners tackle shift issues.

\addition{Next, reproducibility in data science workflows is a well-understood challenge, with attempts to partially address it~\cite{callahan2006vistrails, hull2006taverna, koster2012snakemake, davidson2008provenance}. Recent work also indicates that reproducibility is an ongoing
issue in data science and ML pipelines~\cite{anonymous2021ml, rupprecht2020improving, kery2017variolite, kross2019practitioners}}.
\citet{kross2019practitioners} mention that data science educators who come from industry specifically
want students to learn how to write ``robust and reproducible scientific code.''
In interview studies, \citet{whitherautoml} observe the importance of reproducibility in AutoML workflows,
and \citet{sambasivan2021everyone} mention that practitioners who create reproducible data assets avoid some errors.

Finally, other related work has identified that production ML challenges can be specific to
the ML application at hand. For example, \citet{sambasivan2021everyone} discusses how,
in high-stakes domains like autonomous vehicles, data quality is extra important
and explicitly requires collaboration with domain experts. They explain how data errors
compound and have disastrous impacts, especially in resource-constrained settings.
Unlike the present study, their focus is on data quality issues as opposed to
understanding typical MLE workflows and challenges.
\citet{deploychallenges}
review published reports of individual ML deployments
and mention that not all ML applications can be easily tested prior to deployment.
While ad recommender systems might be easily tested online with a small fraction
of users, other applications require significant simulation testing depending
on safety, security, and scale issues~\cite{rlchallenges,Kuutti2021ASO}. Common applications of ML,
such as medicine~\cite{pumplun2021adoption}, customer service~\cite{folstad2018makes}, and
interview processing~\cite{billington2022machine}
, have their own studies. Our work expands on the literature by identifying common
challenges across various applications and reporting on how MLEs  handle them.

\subsection{Software Engineering for ML}\label{sec:related-sweml}


\addition{Through interviews and practitioner surveys, some papers explore, at a high level,
how ML engineering practices differ from traditional software engineering practices. \citet{Hill2016TrialsAT}
interview ML application developers and report challenges related to building first versions of ML models,
especially around the early stages of exploration and experimentation (e.g., feature engineering, model training).} They describe the process of
building models as ``magic''---similarly echoed by \citet{demystifying} when analyzing ML projects from
Github---with unique practices of debugging data in addition to code. \citet{serbanadoption} conduct a survey of
practitioners and list 29 software engineering practices for ML, such as
``Use Continuous Integration'' and ``Peer Review Training Scripts.'' \citet{muiruri2022practices} interview
Finnish engineers and investigate technical challenges and ML-specific tools in the ML lifecycle.
\citet{amershi_software_2019} identify challenges such as hidden feedback loops and component entanglement
through their interviews with scientists, engineers, and managers at Microsoft. They broadly discuss strategies to integrate
support for ML development into traditional software infrastructure, such as end-to-end pipeline support from data engineers
and educational conferences for employees. Our work expands on the software engineering for ML ecosystem by considering
human-centered, operational requirements for ML deployments, e.g., over time, as MLEs are introduced to ML pipelines
that are unfamiliar to them, or as customer or product requirements change. Unlike Amershi et al.,
we focus on MLEs, who are responsible for maintaining ML pipeline performance. We also interview practitioners
across companies and applications: we provide new and specific examples of ML engineering practices to sustain
ML pipelines as software and categorize these practices around a broader human-centered workflow.

The data management, software engineering, and CSCW communities have proposed various software tools for ML workflows.
For example, some tools manage data provenance and training context for
model debugging purposes~\cite{namaki2020vamsa, vizier, hellerstein2017ground, garcia2020flor}.
Others help ensure reproducibility while iterating on different ideas~~\cite{nbslicer, kery2017variolite, hohman2020understanding}.
With regards to validating changes in \emph{production} systems, some researchers
have studied CI (Continuous Integration) for ML and proposed preliminary solutions---for example,
\texttt{ease.ml/ci} streamlines data management and proposes unit tests for overfitting~\cite{aguilar2021ease}, and
some papers introduce tools to perform validation and monitoring in
production ML pipelines~\cite{dataval, schelter, kang2018model}. Our work is complementary to existing literature on this tooling; we do not explicitly ask interviewees
questions about tools, nor do we propose any tools. We focus on behavioral practices of MLEs.

\addition{\subsection{MLOps Practices and Challenges}~\label{sec:related-mlops}}

\addition{The traditional software engineering literature describes the need for DevOps, a combination of software \emph{dev}elopers and \emph{op}erations teams,
to streamline the process of delivering software in
organizations~\mbox{\cite{leite2019survey, ebert2016devops, lwakatare2019devops, loukides2012devops}}.
Similarly, the field of \emph{MLOps}, or DevOps principles applied to machine learning, has emerged from the rise of
machine learning (ML) application development in software organizations. MLOps is a nascent field, where most existing papers give definitions and overviews of MLOps,
as well as its relation to ML, software engineering, DevOps, and
data engineering~\cite{mlopsoda, mlopsdefs, mlopssurvey, makinen2021needs, tamburri2020sustainable, john2021towards, testi2022mlops, renggli2021data, Shankar2022TowardsOF, Shankar2022RethinkingSM}. Some work in MLOps attempts to characterize a production ML lifecycle; however, there is little consensus.~\citet{mlopsdefs} discuss a lifecycle of data preparation, model selection, and model productionization, but other literature reviews~\cite{lima2022mlops, mlopssurvey} and guides on best practices drawing from authors' experiences~\cite{matsui2022mlops} conclude that, compared to software engineering, there is not yet a standard ML lifecycle, with consensus from researchers and industry professionals. While standardizing an ML lifecycle across different roles (e.g., scientists, researchers, business leaders, engineers) might be challenging, characterizing a workflow specific to a certain role could be more tractable.}

\addition{Several MLOps papers present case studies of productionizing ML within specific organizations and the resulting challenges. For example, adhering to data governance standards and regulation is difficult, as model training is data-hungry by nature~\cite{banerjee2020challenges, granlund2021mlops}.~\citet{mlopssurvey} discuss issues in continuous end-to-end testing (i.e., continuous integration) because ML development involves changes to datasets and model parameters in addition to code. To address such challenges, other MLOps papers have surveyed the proliferating number of industry-originated tools in MLOps~\cite{recupito2022multivocal, mlopsdefs, ruf2021demystifying, lima2022mlops}. MLOps tools can help with general pipeline management, data management, and model management~\cite{recupito2022multivocal}. The surveys on tools motivate understanding how MLEs use such tools, to see if there are any gaps or opportunities for improvement.}

\addition{Prior work in this area---primarily limited to literature reviews, surveys, case studies, and vision papers---motivates research in understanding the {\em human-centered} workflows and pain points in MLOps. Some MLOps work has interviewed people involved in the production ML lifecycle: for example,~\citet{mlopsoda} conduct semi-structured interviews with 8 experts from different industries spanning different roles, such as AI architect and Senior Data Platform Engineer, and uncover a list of MLOps principles such as CI/CD automation, workflow orchestration, and reproducibility, as well as an organizational workflow of product initiation, feature engineering, experimentation, and automated ML workflow pipeline. While~\citet{mlopsoda} explicitly interview professionals from different roles to understand shared patterns between their workflows---in fact, only two of the eight interviewees have ``machine learning engineer'' or ``deep learning engineer'' in their job titles, our work complements their findings by focusing only on self-declared ML engineers responsible for repeated-use models in production and uncovering strategies they use to sustain model performance. As such, we uncover and present a different workflow---one centered around ML engineering. To the best of our knowledge, we are the first to study the human-centered MLOps workflow from ML engineers' perspectives.}




\section{Methods}
\label{sec:methods}

{
\color{revisionBlue}
Upon securing approval from our institution's review board, we conducted an interview study of 18 ML Engineers (MLEs) across various sectors. Our approach mirrored a zigzag model common to Grounded Theory, with alternating phases of fieldwork and in-depth coding and analysis, directing further rounds of interviews~\cite{creswell2016qualitative}. The constant comparative method helped iterate and refine our categories and overarching theory.
Consistent with qualitative research standards, theoretical saturation is generally recognized between 12 to 30 participants, particularly in more uniform populations~\cite{guest2006many}. By our 16th interview, prevalent themes emerged, signaling that saturation was attained. Later interviews confirmed these themes.


Our goal in conducting this study was to develop better tools for ML deployment, broadly targeting monitoring, debugging, and observability issues. Our study was an attempt at identifying key challenges and open opportunities in the MLOps space. This study therefore stems from our collective desire to enrich our understanding of our target community and offer valuable insights into best practices in ML engineering and data science.
Subsequent sections delve into participant recruitment (Subsection~\ref{participant_recruitment}), our interview structure (Subsection~\ref{interview_protocol}), and our coding and analysis techniques (Subsection~\ref{sssec:coding_and_analysis}).
}
\begin{table}[!t]
\color{revisionBlue}
\centering
\resizebox{\textwidth}{!}{%
\begin{tabular}{@{}llllllll@{}}
\toprule
RR & Id & Role         & Org Size & Application       & Yrs Xp & Site   & Highlights                                                                \\ \midrule
1           & Lg1         & MLE Mgr.  & Large        & Autonomous vehicles & 5-10     & US-West& high velocity experimentation; scenario testing                                   \\
1           & Md1         & MLE          & Medium       & Autonomous vehicles & 5-10     & US-West& pipeline-on-a-schedule; copy-paste anomalies                                    \\
1           & Sm1         & MLE          & Small        & Computer hardware   & 10-15    & US-West& exploratory data analysis; AB Testing; SLOs                                     \\
1           & Md2         & MLE          & Medium       & Retail              & 5-10     & US-East& retraining cadence; adaptive test data; feedback delay                           \\
1           & Lg2         & MLE Mgr.  & Large        & Ads                 & 5-10     & US-West& ad click count; model ownership; keeping GPUs warm                               \\
1           & Lg3         & MLE          & Large        & Cloud computing     & 10-15    & US-West& bucketing / binning; SLOs; hourly batched predictions                            \\
2           & Sm2         & MLE          & Small        & Finance             & 5-10     & US-West& F1-score ; retraining cadence; progressive validation                             \\
2           & Sm3         & MLE          & Small        & NLP                 & 10-15    & Intl   & triage queue; fallback models; false-positive rate                                 \\
2           & Sm4         & MLE          & Small        & OCR + NLP           & 5-10     & Intl   & human annotators; word2vec; airflow                                               \\
3           & Md3         & MLE Mgr.  & Medium       & Banking             & 10-15    & US-West& human annotators; institutional knowledge; revenue                                \\
3           & Lg4         & MLE          & Large        & Cloud computing     & 10-15    & US-West& online inference; pipeline-on-a-schedule; fallback models                          \\
3           & Sm5         & MLE          & Small        & Bioinformatics      & 5-10     & US-West& model fine-tuning; someone else's features                                        \\
4           & Md4         & MLE          & Medium       & Cybersecurity       & 10-15    & US-East& model-per-customer; join predictions w/ ground truth                               \\
4           & Md5         & MLE          & Medium       & Fintech             & 10-15    & US-West& retraining cadence; dropped special characters            \\
5           & Sm6         & MLE          & Small        & Marketing and analytics & 5-10 & US-East& human annotators; label quality; adaptive test data                              \\
5           & Md6         & MLE          & Medium       & Website builder     & 5-10     & US-East& SLOs; poor documentation; data validation                \\
6           & Lg5         & MLE          & Large        & Recommender systems & 10-15    & US-West& pipeline-on-a-schedule; SLOs; progressive validation   \\
6           & Lg6         & MLE Mgr.  & Large        & Ads                 & 10-15    & US-West& fallback models; on-call rotations; scale                 \\ \bottomrule
\end{tabular}%
}
\caption{\color{revisionBlue} The table provides an anonymized description of interviewees from different sizes of companies, roles, years of experience, application areas, and their code attributions. The interviewees hail from a diverse set of backgrounds. \additiontwo{Small companies
have fewer than 100 employees; medium-sized companies have 100 -
1000 employees, and large companies have 1000 or more employees.} \textbf{RR} denotes recruitment round. \textbf{Highlights} refers to key codes (i.e. from the code system in \cref{fig:codebook}) attached to that participant's transcript.}
\label{tab:interviewees}
\end{table}

\subsection{Participant Recruitment \& Selection}\label{participant_recruitment}

We recruited individuals
who were responsible for the development, regular retraining, monitoring and deployment of ML models \emph{in production}.
A description of the 18 MLEs is shown in~\Cref{tab:interviewees}.
The MLEs interviewed varied in their educational backgrounds, years of experience, roles, team size, and work sector.
Recruitment was conducted in rounds over the course of an academic year (2021-2022).
{
\color{revisionBlue}
Our recruitment strategy was underpinned by a deliberate, iterative process that built upon the insights from each round. The primary goal was to cultivate a representative sample that captured the rich diversity of Machine Learning Engineers (MLEs) across various dimensions.

\subsubsection{Initial Recruitment: Relying on Professional Networks}
\label{sssec:initial_recruitment}

In the initial recruitment round (RR=1), we leaned heavily on the established professional networks of our faculty co-authors. This approach, while convenient and efficient, resulted in a sample that was geographically skewed towards the US-West. It also led to a greater representation from larger organizations, as well as certain sectors like Autonomous Vehicles and Cloud Computing. This initial cohort provided valuable insights but also highlighted the potential biases and gaps in our sample.

\subsubsection{Course Correction: Diversifying the Sample}
\label{sssec:course_correction}

Recognizing the need for a more representative and diversified sample, our strategy in subsequent rounds shifted. Specifically for RR=2, we made a concerted effort to engage candidates outside our immediate professional networks and particularly targeted those at smaller companies. This shift in approach was operationalized by posting recruitment advertisements and flyers in various online communities. 
Prospective participants who responded to our outreach underwent a screening process for the same inclusion criteria mentioned previously. Their professional backgrounds and affiliations were verified through platforms such as professional websites, LinkedIn, and online portfolios. As a result, we observed a stronger representation from domains like NLP and Finance.

\subsubsection{Building a Representative Sample: Iterative Refinement}
\label{sssec:iterative_refinement}

Each recruitment round served as a feedback loop, informing the strategy for the subsequent round. As patterns emerged from our data analysis, we fine-tuned our recruitment focus to fill identified gaps. This iterative process ensured that, over time, our sample grew to be more balanced in terms of roles, experience, organization sizes, sectors, and geographical locations. By employing this iterative recruitment strategy, we believe our study encapsulates a comprehensive cross-section of the MLE community, offering insights that are both deep and broad.
}

In each round, between three to five candidates were reached by email and invited to participate.
We relied on our professional networks and open calls posted on MLOps channels in Discord\footnote{https://discord.com/invite/Mw77HPrgjF}, Slack\footnote{mlops-community.slack.com}, and Twitter to compile a roster of candidates.
The roster was incrementally updated roughly after every round of interviews, integrating information gained from the concurrent coding and analysis of transcripts (Section~\ref{sssec:coding_and_analysis}).
Recruitment rounds were repeated until
we reached saturation on our findings~\cite{muller2014curiosity}.

\subsection{Interview Protocol}\label{interview_protocol}

With each participant, we
conducted semi-structured interviews
over video call lasting 45 to 75 minutes each.
Over the course of the interview, we asked descriptive,
structural, and contrast questions abiding by ethnographic interview guidelines~\cite{spradley2016ethnographic}.
The questions are listed in \Cref{app:interviewq}.
Specifically, our questions spanned six categories:
i) the type of ML tasks they work on,
ii) the approaches used for building or tuning models,
iii) the transition from development to production,
iv) how they evaluate their models before deployment,
v) how they monitor their deployed models, and
vi) how they respond to issues or bugs.
Participants received and signed a consent form before the interview, and agreed to participate free of compensation.
As per our agreement, we automatically transcribed the interviews using Zoom software.
In the interest of privacy and confidentiality, we did not record audio or video of the interviews.
Transcripts were redacted of personally identifiable information before being uploaded to a secured drive in the cloud.

\subsection{Transcript Coding \& Analysis}\label{sssec:coding_and_analysis}

We employed grounded theory to systematically analyze our interview transcripts. Grounded theory is a robust methodology focused on iterative data collection and analysis for theory discovery~\cite{strauss1994grounded,cho2014reducing}. One of its key features is the seamless integration of data collection and analysis, aiming to identify emerging themes and concepts through a constant review of transcripts. \additiontwo{In employing grounded theory, we followed its key processes: open, axial, and selective coding. During {\em open coding}, the initial phase of categorizing data, we deconstructed our interview transcripts into discrete ideas or phenomena and assigned codes to these ideas (e.g., \textit{A/B testing}). Then, in {\em axial coding}, where the goal is to identify patterns and relationships between different concepts, we merged duplicate codes and drew edges between similar codes. For example, we grouped the codes \textit{scenario testing} and \textit{A/B testing} under the broader \textit{testing} code. Finally, through {\em selective coding}, we distilled our codes into five core themes that represent the essence of our transcripts. \Cref{fig:codebook} shows our hierarchy of codes, with core themes such as \textbf{Tasks}, \textbf{Operations}, and \textbf{Systems \& Tools}.}
As illustrated in~\Cref{fig:visualoverview}, we allocated roughly equal time to each main theme, which correspondingly informed our findings. The themes relate to our findings as follows:

\begin{figure*}
\centering
\begin{subfigure}[t]{0.64\textwidth}
\includegraphics[width=\textwidth]{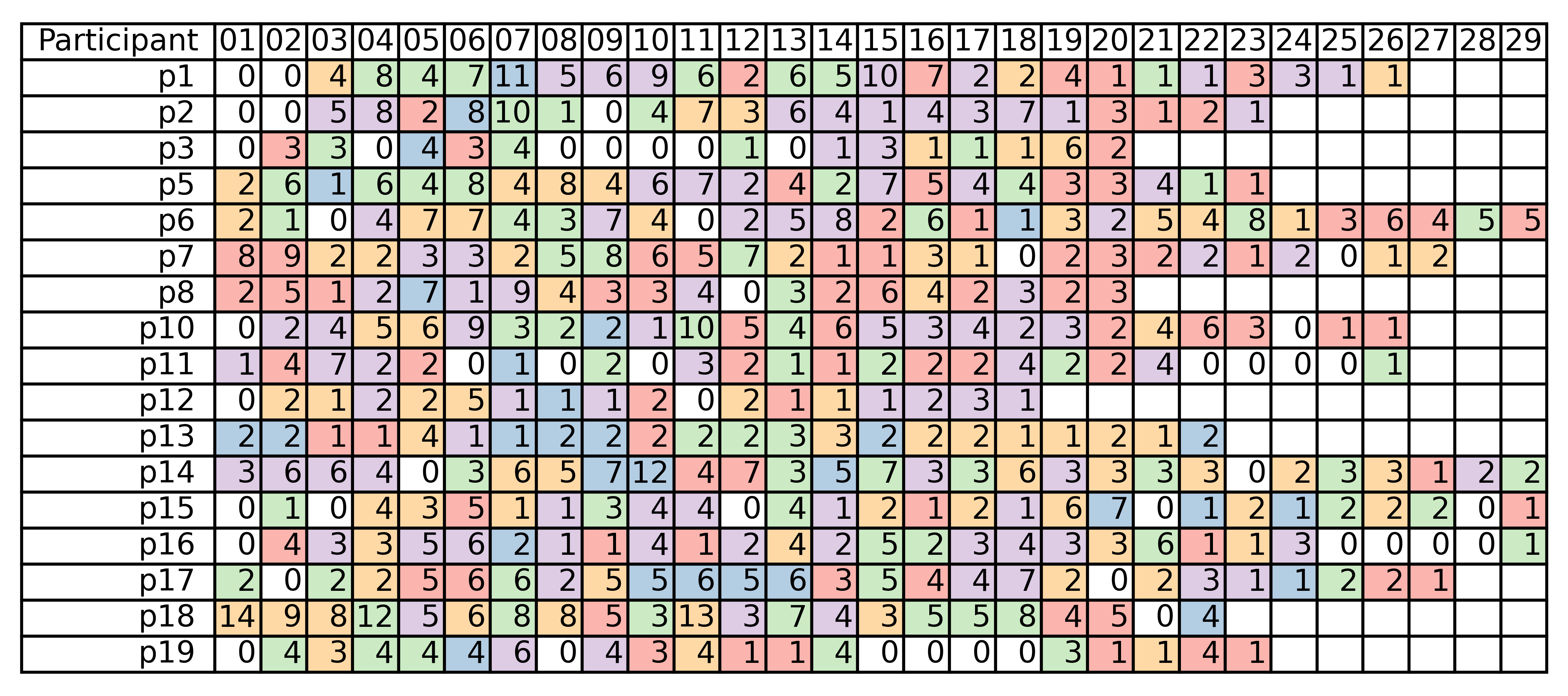}
\label{fig:segmentColors}
  \caption{Color-coded Overview of Transcripts}
\end{subfigure}
\hfill
\begin{subfigure}[t]{0.33\textwidth}
\includegraphics[width=\textwidth]{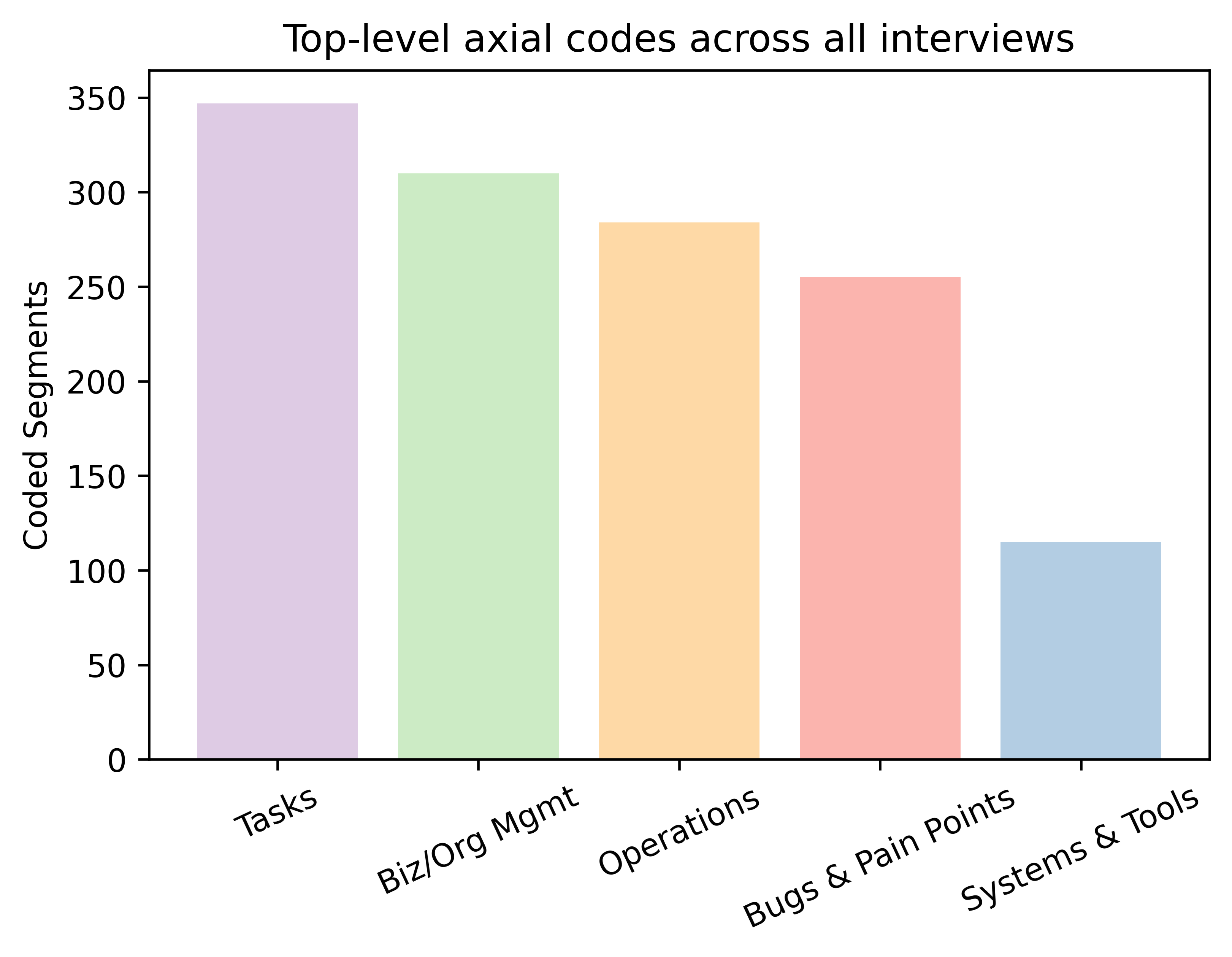}
\label{fig:colorHistogram}
  \caption{Color Legend}
\end{subfigure}
 \caption{Visual summary of coded transcripts. The x-axis of (a), the color-coded overview, corresponds to a segment (or group) of transcript lines, and the number in each cell is the code's frequency for that transcript segment and for that participant. Segments are blank after the conclusion of each interview, and different interviews had different time duration. Each color in (a) is associated with a top-level axial code from our interview study, and presented in the color legend (b). The color legend also shows the frequency of each code across all interviews.}
\label{fig:visualoverview}
\end{figure*}

\begin{figure}[!t]
\color{revisionBlue}
\centering
\begin{enumerate}
    \item \textbf{Tasks}
    \begin{enumerate}
        \item Data collection, cleaning \& labeling: \textit{human annotators}, \textit{exploratory data analysis}
        \item Embeddings \& feature engineering: \textit{normalization}, \textit{bucketing / binning}, \textit{word2vec}
        \item Data modeling \& experimentation: \textit{accuracy}, \textit{F1-score}, \textit{precision}, \textit{recall}
        \item Testing: \textit{scenario testing}, \textit{AB testing}, \textit{adaptive test-data}
    \end{enumerate}
    \item \textbf{Biz/Org Management}
    \begin{enumerate}
        \item Business focus: \textit{service level objectives}, \textit{ad click count}, \textit{revenue}
        \item Teams \& collaboration: \textit{institutional knowledge}, \textit{on-call rotations}, \textit{model ownership}
        \item Process maturity indicators: \textit{pipeline-on-a-schedule},  \textit{fallback models}, \textit{model-per-customer}
    \end{enumerate}
    \item \textbf{Operations}
    \begin{enumerate}
        \item CI/CD: \textit{artifact versioning}, \textit{multi-staged deployment}, \textit{progressive validation}
        \item Data ingestion \& integration: \textit{automated featurization}, \textit{data validation} 
        \item Model retraining: \textit{distributed training}, \textit{retraining cadence}, \textit{model fine-tuning}
        \item Prediction serving: \textit{hourly batched predictions}, \textit{online inference}
        \item Live monitoring: \textit{false-positive rate}, \textit{join predictions w/ ground truth}
    \end{enumerate}
    \item \textbf{Bugs \& Pain Points}
    \begin{enumerate}
        \item Bugs: \textit{data leakage}, \textit{dropped special characters}, \textit{copy-paste anomalies}
        \item Pain points: \textit{big org red tape}, \textit{performance regressions}, \textit{label quality}, \textit{scale}
        \item Anti-patterns: \textit{muting alerts}, \textit{keeping GPUs warm}, \textit{waiting it out}
        \item Known challenges: \textit{data drift}, \textit{feedback delay}, \textit{class imbalance}, \textit{sensor divergence}
        \item Missing context: \textit{someone else's features}, \textit{poor documentation}, \textit{much time has passed}
    \end{enumerate}
    \item \textbf{Systems \& Tools}
    \begin{enumerate}
        \item Metadata layer: \textit{Huggingface},  \textit{Weights \& Biases}, \textit{MLFlow}, \textit{TensorBoard}, \textit{DVC}
        \item Unit layer: \textit{PyTorch}, \textit{TensorFlow}, \textit{Jupyter}, \textit{Spark}
        \item Pipeline layer: \textit{Airflow}, \textit{Kubeflow}, \textit{Papermill}, \textit{DBT}, \textit{Vertex AI}
        \item Infrastructure layer: \textit{Slurm}, \textit{S3}, \textit{EC2}, \textit{GCP}, \textit{HDFS}
    \end{enumerate}
\end{enumerate}
\caption{\color{revisionBlue} Abridged code system: A distilled representation of the evolved code system resulting from our qualitative study, capturing the primary tasks, organizational aspects, operational methodologies, challenges, and tools utilized by Machine Learning Engineers.}
\label{fig:codebook}
\end{figure}


\begin{enumerate}
      \item \textbf{Tasks} refers to activities that are routinely performed by ML engineers.
            The analysis of code segments descended from \textit{tasks},
            and its decomposition into constituent parts, culminated in the creation of
            the MLOps workflow (\Cref{fig:tasksdiagram}), and is instrumental in the structure and presentation of~\cref{sec:summary} (\emph{Findings}).

      \item \textbf{Biz/Org (Business-Organizational) Management} refers to modes of interaction between MLEs and their co-workers or managers, and between MLEs and customers or other stakeholders.
            Relevant sub-codes form the theoretical basis for~\Cref{sec:collab} (\emph{Collaboration})
            and \Cref{sec:summary-eval-product} (\emph{Product Metrics}).

      \item \textbf{Operations} refers to repeatable work that must be performed regularly and consistently for the continued operation of the business.
            \textit{Operations} is the ``Ops'' in MLOps.
            Relevant sub-codes form the theoretical basis for~\Cref{sec:dag} (\emph{Pipeline Automation}).

      \item \textbf{Bugs \& Pain Points} refers to failure modes encountered at any stage in the MLOps workflows, MLE complaints generally, and author-noted anti-patterns. These are discussed in Monitoring and Response (\Cref{sec:summary-monitoring}).
      
      \item \textbf{Systems \& Tools} refers to storage and compute infrastructure, programming languages,
            ML training frameworks, experiment execution frameworks, and other tools or systems that MLEs use in their MLOps work. We discuss implications for tool design in~\Cref{sec:discussion-tools}. \addition{We include a table of common tools referenced by interviewees in~\Cref{app:tools}.}

\end{enumerate}

\section{Summary of Findings}
\label{sec:summary}

Going into the interview study, we assumed the \addition{workflow of human-centered tasks in the production ML lifecycle was similar to the production data science workflow presented by \citet{autoai},
which is a loop consisting of the following:}

\begin{enumerate}
    \item \textbf{Preparation}, spanning data acquisition, data cleaning and labeling, and feature engineering,
    \item \textbf{Modeling}, spanning model selection, hyperparameter tuning, and model validation, and
    \item \textbf{Deployment}, spanning model deployment, runtime monitoring, and model improvement.
\end{enumerate}

\addition{From our interviews, we found that the repeated-use production ML workflow that ML engineers engage in differs slightly from \citet{autoai}. As preliminary research papers defining and providing reference architectures for MLOps have pointed out, operationalizing ML brings new requirements to the table, such as the need for teams,
not just individual people, to understand, sustain, and improve ML pipelines and systems over time~\cite{mlopsoda, mlopsdefs, mlopssurvey}.
In the pipelines that our interviewees build and supervise, most technical components are
automated---e.g., data preprocessing jobs run on a schedule, and models
are typically retrained regularly on fresh data.
We found the ML engineering workflow to revolve around the following stages (\Cref{fig:tasksdiagram}):}

\begin{enumerate}
    \item \textbf{Data Preparation}, which includes scheduled data acquisition, cleaning, labeling, and transformation,
    \item \textbf{Experimentation}, which includes both data-driven and model-driven changes to increase overall ML performance,
          and is typically measured by metrics such as accuracy or mean-squared-error,
    \item \textbf{Evaluation \& Deployment}, which consists of a model retraining loop which periodically rolls out a new model---or
          offline batched predictions, or another proposed change---to growing fractions of the population, and
    \item \textbf{Monitoring \& Response}, which supports the other stages via data and code instrumentation
          (e.g., tracking available GPUs for experimentation or the fraction of null-valued data points) and dispatches engineers and bug fixes to identified problems in the pipeline.
\end{enumerate}

For each stage, we identified human-centered practices from the \emph{Tasks},
\emph{Biz/Org Management}, and \emph{Bugs \& Pain Points} codes,
and drew on \emph{Operations} codes for automated practices
(refer to \Cref{sssec:coding_and_analysis} for a description of these codes).
\addition{An overview of findings for each workflow stage can be found in \Cref{tab:summaryfindings}.} 

\begin{table}[!t]
    \centering
    \small 
    \addition{\begin{tabular}{p{2.25cm}p{4.5cm}p{6cm}}
        \toprule
        \textbf{Stage} & \textbf{Description} & \textbf{Non-Automated Challenges} \\
        \midrule
        \multirow{2}{*}{Data Preparation} & \multirow{2}{4.5cm}{Collection, wrangling, and cleaning pipelines run on a schedule} & - Ensuring label quality at scale (\Cref{sec:summary-dataprep-labelquality}) \\
        & & - Handling feedback or ground-truth delays  (\Cref{sec:summary-dataprep-feedbackdelays}) \\
        \midrule
        \multirow{3}{*}{Experimentation} & \multirow{3}{4.5cm}{Prototyping ideas to improve end-to-end ML pipeline performance by iterating on datasets, model architectures, or both} & - Managing the underlying software or code for data-centric experiments (\Cref{sec:summary-expt-datacentric}) \\
        & & - Engaging in cross-team collaboration (\Cref{sec:collab}) \\
        & & - Manually and thoughtfully identifying promising experiment configurations (\Cref{sec:summary-expt-iterative}) \\
        \midrule
        \multirow{2}{2.25cm}{Evaluation and Deployment} & \multirow{2}{4.5cm}{Pushing experimental changes to small, then large fractions of users, evaluating at every step} & - Continuously updating dynamic validation datasets for future experiments (\Cref{sec:summary-eval-dynamic}) \\
        & & - Using product metrics for evaluation (\Cref{sec:summary-eval-product}) \\
        \midrule
        \multirow{3}{2.25cm}{Monitoring and Response} & \multirow{3}{4.5cm}{Supervising live ML pipeline performance and minimizing pipeline downtime} & - Tracking and investigating data quality alerts (\Cref{sec:summary-hacks-dataval}) \\
        & & - Managing pipeline "jungles" of models and hard-coded rules (\Cref{sec:summary-monitoring-jungle}) \\
        & & - Debugging a heavy-tailed distribution of errors (\Cref{sec:summary-monitoring-bugs}) \\
        \bottomrule
    \end{tabular}}
    \caption{\addition{Overview of challenging activities that ML engineers engage in for each stage in their workflow. While each stage relies on automated infrastructure and pipelines, ML engineers still have many difficult manual responsibilities.}}
    \label{tab:summaryfindings}
\end{table}

The following subsections organize our findings around the four stages of MLOps.
We begin each subsection with a quote that stood out to us and conversation
with prior work; then, in the context of what is automated,
we discuss common human-centered practices and pain points.

\subsection{Data Preparation}\label{sec:summary-dataprep}

\begin{displayquote}
    ``It takes exponentially more data to keep getting linear increases in performance.'' --\addition{Lg1}
\end{displayquote}

Data preparation is the process of constructing ``well-structured, complete datasets'' for data scientists~\cite{muller2019data}.
Data preparation activities consist of collection, wrangling, and cleaning and are known to be challenging, often taking up to 80\% of practitioners' time~\cite{kandel2011wrangler, autoai}.
This tedious process encourages
larger organizations to have dedicated teams of data engineers to manage
data preparation~\cite{kandel}. Like \citet{amershi_software_2019}, we observed that mature ML
organizations automated data preparation through dedicated teams
as much as possible (\addition{Lg1}, \addition{Lg2}, \addition{Lg3}, \addition{Sm3}, \addition{Md3}, \addition{Sm6}, \addition{Md6}, \addition{Lg5}, \addition{Lg6}). As a result, the MLEs we interviewed spent a smaller fraction of their time
on data preparation, collaborating instead with data engineering teams.
We first discuss pipeline automation to provide key context for the work of MLEs.
Then, we mention two key challenges MLEs face: ensuring labeling quality at scale and coping with feedback delays.

\subsubsection{\textbf{Pipelines automatically run on a schedule}}\label{sec:dag}

Unlike data science, where data preparation is often ad-hoc and interactive~\cite{kandel, muller2019data}, we found that
data preparation in production ML is batched and more narrowly restricted,
involving an organization-wide set of steps running at a predefined cadence.
In interviews, we found that preparation pipelines were defined by Directed Acyclic Graphs, or DAGs, which ran on a schedule (e.g., daily).
Each DAG node corresponded to a particular task, such as ingesting data from a source or cleaning a newly ingested partition of data.
Each DAG edge corresponded to a dataflow dependency between tasks. While data engineers were primarily responsible for the end-to-end
execution of data preparation DAGs, MLEs interfaced with these DAGs by loading select outputs (e.g., clean data) or by extending the
DAG with additional tasks, e.g. to compute new features (\addition{Md1}, \addition{Lg2}, \addition{Lg3}, \addition{Sm4}, \addition{Md6}, \addition{Lg6}).

In many cases, automated tasks relating to \emph{ML models}, such as model inference (e.g., generating predictions with a trained model)
and retraining, were executed in the same DAG as data preparation tasks (\addition{Lg1}, \addition{Md1}, \addition{Sm2}, \addition{Md4}, \addition{Md5}, \addition{Sm6}, \addition{Md6}, \addition{Lg5}). ML engineers included
retraining as a node in the data preparation DAG for simplicity: as new data becomes available, a corresponding model is automatically retrained.
\addition{Md4} mentioned automatically retraining the model every day so model performance would not suffer for longer than a day:

\begin{displayquote}
    Why did we start training daily? As far as I'm aware, we wanted to start simple---we could
    just have a single batch job that processes new data and we wouldn't need to worry about
    separate retraining schedules. You don't really need to worry about if your model has gone
    stale if you're retraining it every day.
\end{displayquote}

Retraining cadences ranged from hourly (\addition{Lg5}) to every few months (\addition{Md6}) and
were different for different models within the same organization (\addition{Lg1}, \addition{Md4}). \addition{None of the
participants interviewed reported any scientific procedure for determining the pipeline execution cadence. For example, \addition{Md5} said that \blockquote{the [model retraining] cadence was just like, finger to the wind.} Cadences seemed to be set in a way that streamlined operations for the organization in the easiest way.}
\addition{Lg5} mentioned that
\blockquote{retraining took about 3 to 4 hours, so [they]
    matched the cadence with it such that as soon as [they] finished any one model, they kicked off the next training [job].}
Engineers reported an inability to retrain unless they had fresh and labeled data,
motivating their organizations to set up dedicated teams of annotators, or hiring crowd workers,
to operationalize labeling of live data (\addition{Lg1}, \addition{Sm3}, \addition{Sm4}, \addition{Md3}, \addition{Sm6}).

\subsubsection{\textbf{MLEs ensure label quality at scale}}
\label{sec:summary-dataprep-labelquality}

Although it is widely recognized that model performance improves with more labels~\cite{roh_survey_2019}---and
there are tools built especially for data labeling~\cite{ratner2017snorkel, zhang2022onelabeler}---our
interviewees cautioned that the quality of labels can degrade as they try to label more and more data. \addition{Md3} said:

\begin{displayquote}
    No matter how many labels you generate, you need to know what you're actually labeling and you need to have a very clear human definition of what they mean.
\end{displayquote}

In many cases, ground truth must be \emph{created}, i.e., the labels are what a practitioner ``thinks up''~\cite{muller2019data}.
When operationalizing this practice, MLEs faced problems. For one, \addition{Sm3} spoke about how expensive it was to outsource labeling.
Moreover, labeling conflicts can erode trust in data quality, and slow ML progress~\cite{passi2018trust,sambasivan2021everyone}:
When scaling up labeling---through labeling service providers or analysts within the organization---MLEs frequently found disagreements
between different labelers, which would negatively impact quality if unresolved (\addition{Sm3}, \addition{Md3}, \addition{Sm6}).
\addition{At their organization, \addition{Sm3} mentioned that there was a human-in-the-loop labeling pipeline that both outsourced
large-scale labeling and maintained an internal team of experts to verify the labels and resolve errors manually.} \addition{Sm6} described a label validation pipeline for outsourced labels that itself used ML models for estimating label quality.

\subsubsection{\textbf{Feedback delays can disrupt pipeline cadences}}\label{sec:summary-dataprep-feedbackdelays}
In many applications, today's predictions are tomorrow's training data,
but many participants said that ground-truth labels for live predictions often arrived after a delay,
which could vary unpredictably (e.g., human-in-the-loop or networking delays) and thus
caused problems for knowing real-time performance or retraining regularly (\addition{Md1}, \addition{Sm2}, \addition{Sm3}, \addition{Md5}, \addition{Md6}, \addition{Lg5}).
This is in contrast to academic ML, where ground-truth labels are readily available for ML
practitioners to train models~\cite{deploychallenges}. Participants noted that the impact on
models was hard to assess when the ground truth involved live data---for example,
\addition{Sm2} felt strongly about the negative impact of feedback delays on their ML pipelines:

\begin{displayquote}
    I have no idea how well [models] actually perform on live data. Feedback is always delayed by at least 2 weeks. Sometimes we
    might not have feedback...so when we realize maybe something went wrong, it could [have been] 2 weeks ago.
\end{displayquote}

Feedback delays contribute to ``data cascades,'' or compounding errors in ML systems over time~\cite{sambasivan2021everyone}.
\addition{Sm3} mentioned a 2-3 \emph{year} effort to develop a human-in-the-loop pipeline to manually label live
data as frequently as possible to side-step feedback delays:
\blockquote{you want to come up with the rate at which data is changing, and then assign people to manage this rate roughly}.
\addition{Sm6} said that their organization hired freelancers to label ``20 or so'' data points by hand daily.
Labeling was then considered a task in the broader preparation pipeline that ran on a schedule (\Cref{sec:dag}).

\subsection{Experimentation}\label{sec:summary-expt}
\begin{displayquote}
    ``You want to see some degree of experimental thoroughness. People will have principled stances or
    intuitions for why things should work. But the most important thing to do is achieve scary high
    experimentation velocity...Number one [Key Performance Indicator] is rate of experimentation.'' (\addition{Lg1}
\end{displayquote}

\addition{While most prior work studying the data science and MLOps workflows includes \emph{modeling} as an explicit step in the workflow~\cite{autoai,zhang2020data,Wongsuphasawat2019GoalsPA,studer2021towards},
we found that iterating on model ideas and architectures is only part of a broader ``experimentation'' step.} This is because in many
production ML pipelines, MLEs can focus on tuning or improving existing models \addition{through data-centric development}, and \emph{modeling} in
a data science sense is only necessary when the company wishes to expand its service offerings or grow its ML capabilities. \addition{In fact, many of our interviewees did not build the initial model in the pipeline that their organization assigned them to work on, so their goal wasn't necessarily to train more models. As an example, \addition{Md6} said, ``some of our models have been around for, like, 6 or 7 years.''} \addition{\citet{mlopssurvey} also call this workflow step ``experimentation'' instead of ``modeling'' in their MLOps lifecycle overview, and we expand on this finding in our paper by relating it to collaboration and data-driven exploration, as well as MLE reservations toward experiment automation or AutoML.}

\subsubsection{\textbf{MLEs find it better to innovate on the data than the model}}\label{sec:summary-expt-datacentric}
\citet{holstein_improving_2019} mention that it is challenging for practitioners to determine where to focus experimentation
efforts---they could try ``switching to a different model, augmenting the training data in some way,
collecting more or different kinds of data, post-processing outputs, changing the objective function, or something else.''
Our interviewees recommended focusing on experiments that provided additional context to the model, typically via new features,
to get the biggest gains
(\addition{Lg2}, \addition{Lg3}, \addition{Md3}, \addition{Lg4}, \addition{Md4}, \addition{Sm6}, \addition{Md6}, \addition{Lg5}, \addition{Lg6}). \addition{Lg5} said:

\begin{displayquote}
    I'm focusing my energy these days on signals and feature engineering because even if you keep your code static and the model static, it would definitely help you with getting better model performance.
\end{displayquote}

In a concurring view, \addition{Md3} adds:

\begin{displayquote}
    I'm gonna start with a [fixed] model because it means faster iterations. And often---like most of the time empirically---it's
    going to be something in our data that we can use to push the boundary [\dots] \addition{obviously}, it's not a dogmatic ``we will never touch the model'' but it shouldn't be our first move.
\end{displayquote}

Interestingly, older work claims that iterating on the model is often more fruitful than iterating
on the data~\cite{Patel2008InvestigatingSM}, but this could be because ML modeling libraries weren't
as mature as they are now. Recent work has also identified the importance of data-centric experimentation
in production ML deployments~\cite{sambasivan2021everyone, deploychallenges, muller2019data,ratner2017snorkel}.
\addition{Md6} mentioned that most ML projects at their organization centered around adding new features.
\addition{Md4} mentioned that one of their current projects was to move feature engineering pipelines from Scala to SparkSQL
(a language more familiar to ML engineers and data scientists), so experiment ideas could be coded and validated faster.

When asked how they managed the underlying software or code for data-centric experiments,
interviewees emphasized the importance of keeping their code changes as small as possible for multiple reasons,
including faster code review, easier validation, and fewer merge conflicts (\addition{Md1}, \addition{Lg2}, \addition{Lg3}, \addition{Sm4}, \addition{Md3}, \addition{Lg5}, \addition{Lg6}). \
This is in line with good software engineering practices~\cite{amershi_software_2019}.
Additionally, changes in large organizations were primarily made in configuration (config) files instead of main
application code (\addition{Lg1}, \addition{Md1}, \addition{Lg2}, \addition{Sm4}, \addition{Lg4}, \addition{Lg6}): instead of editing parameters directly in a Python model training script,
MLEs preferred to edit a config file of parameters (e.g., JSON or YAML), and would feed the config file to the
model training script.
When larger changes were necessary, especially changes touching the language layer
(e.g. changing PyTorch or TensorFlow architectures), MLEs would fork the code base and made their edits
in-place (\addition{Md2}, \addition{Lg3}). Although forking repositories can be a high-velocity shortcut, absent streamlined merge procedures,
this can lead to a divergence in versions and accumulation of technical debt. \addition{Lg3} highlighted the tension
between experiment velocity and strict software engineering discipline:

\begin{displayquote}
    I used to see a lot of people complaining that model developers don't follow software engineering.
    At this point, I'm feeling more convinced that it's not because they're lazy.
    It's because [software engineering is] contradictory to the agility of analysis and exploration.
\end{displayquote}

\subsubsection{\textbf{Feature engineering is social and collaborative}}\label{sec:collab}
Prior work has stressed the importance of collaboration in data science projects,
often lamenting that technical tasks happen in silos~\cite{krossorienting,sambasivan2021everyone,autoai,deploychallenges}.
Our interviewees similarly believed cross-team collaboration was critical for successful experiments.
Project ideas, such as new features, came from or were validated early by domain experts:
data scientists and analysts who had already performed a lot of exploratory data analysis.
\addition{Md4} and \addition{Md6} independently recounted successful project ideas that came from asynchronous conversations on Slack:
\addition{Md6} said, \blockquote{I look for features from data scientists, [who have ideas of] things that are correlated with what I'm trying to predict.}
We found that organizations explicitly prioritized cross-team collaboration as part of their ML culture. \addition{Md3} said:

\begin{displayquote}
    We really think it's important to bridge that gap between what's often, you know, a [subject matter expert] in one room annotating and then handing things over the
    wire to a data scientist---a scene where you have no communication. So we make sure there's both data science and subject matter expertise representation [in our meetings].
\end{displayquote}

To foster a more collaborative culture, \addition{Sm6} discussed the concept of ``building goodwill'' with
other teams through tedious tasks that weren't always explicitly a part of company plans:
\blockquote{Sometimes we'll fix something [here and] there to like build some goodwill, so that we can call on them in the future...I do this stuff [to build relationships],
    not because I'm really passionate about doing it.}

\subsubsection{\textbf{MLEs like having manual control over experiment selection}}\label{sec:summary-expt-iterative}
One challenge that results from fast exploration is having to manage many experiment versions~\cite{autoai, mlopsoda}.
MLEs are happy to delegate experiment tracking and execution work to ML experiment execution frameworks,
such as Weights \& Biases\footnote{\addition{https://wandb.ai/}}, but prefer to choose subsequent experiments themselves.
To be able to make informed choices of subsequent experiments to run, MLEs must maintain awareness
of what they have tried and what they haven't (\addition{Lg2} calls it the ``exploration frontier'').
As a result, there are limits to how much automation MLEs are willing to rely on for experimentation,
a finding consistent with results from~\citet{whitherautoml}.
\addition{Lg2} mentioned the phrase \blockquote{keeping GPUs warm} to characterize a perceived anti-pattern that wastes resources without a plan:

\begin{displayquote}
    One thing that I've noticed is, especially when you have as many resources as [large companies] do,
    that there's a compulsive need to leverage all the resources that you have.
    And just, you know, get all the experiments out there.
    Come up with a bunch of ideas; run a bunch of stuff. I actually think that's bad.
    You can be overly concerned with keeping your GPUs warm, [so much] so that you don't actually think deeply about what the highest value experiment is.
\end{displayquote}

In executing experiment ideas, we noticed a tradeoff between a guided search and random search.
Random searches were more suited to parallelization---e.g.,
hyperparameter searches or ideas that didn't depend on each other.
Although computing infrastructure could support many different experiments in parallel,
the cognitive load of managing such experiments was too cumbersome for participants (\addition{Lg2}, \addition{Sm4}, \addition{Lg5}, \addition{Lg6}).
Rather, participants noted more success when pipelining learnings from one experiment into the next,
like a guided search to find the best idea (\addition{Lg2}, \addition{Sm4}, \addition{Lg5}).  \addition{Lg5} described their ideological shift from random search to guided search:

\begin{displayquote}
    Previously, I tried to do a lot of parallelization. If I focus on one idea, a week at a time, then it boosts my productivity a lot more.
\end{displayquote}

By following a guided search, engineers are, essentially, significantly pruning a large subset of experiment ideas without executing them.
While it may seem like there are unlimited computational resources, the search space is much larger, and developer time and energy is limited. At the end of the day,
experiments are human-validated and deployed. Mature ML engineers know their personal tradeoff between parallelizing disjoint
experiment ideas and pipelining ideas that build on top of each other, ultimately yielding successful deployments.

\subsection{Evaluation and Deployment}\label{sec:summary-eval}

\begin{displayquote}
    ``We don't have a good idea of how the model is going to behave in production until production.'' (\addition{Lg3})
\end{displayquote}

After the experimentation phase,
when MLEs have identified a change they want to make to the ML pipeline
(e.g., adding a new feature), they need to evaluate it and deploy it to production. \addition{Prior work that prescribes frameworks for MLOps typically separates evaluation and deployment into two different stages~\cite{mlopsdefs, studer2021towards, kumara2022requirements}, but we combine them into one step because they are tightly intertwined, with deployments spanning long periods of time and evaluations happening multiple times during deployment.}

Prior work describes evaluation as an ``offline,'' automated process that happens
at training time: a small portion of the training dataset is held out, and the model
should achieve high accuracy on this held-out set~\cite{deploychallenges, Wongsuphasawat2019GoalsPA}. \addition{Recent related work in MLOps claims that evaluation and deployment are highly amenable to automation~\cite{mlopssurvey, matsui2022mlops}. As such, we also originally hypothesized that evaluation and deployment could be automated---once
validated, an engineer could simply create a new task in their DAG to retrain the model on a cadence (\Cref{sec:dag}).}

As expected, engineers did automatically validate and codify their changes
into DAGs to retrain models on a schedule. \addition{However, they also manually supervised the deployment over a
long period of time, evaluating throughout the time frame.} \citet{amershi_software_2019} state that software teams
``flight'' changes or updates to ML models, often by testing them on a few cases prior to live deployment.
Our work provides further context into the evaluation and deployment process for production ML pipelines:
we found that several organizations, particularly those with many customers,
employed a \emph{multi-stage deployment process} for new models or model changes,
progressively evaluating at each stage (\addition{Sm1}, \addition{Lg2}, \addition{Lg3}, \addition{Sm2}, \addition{Sm3}, \addition{Lg4}, \addition{Md5}, \addition{Md6}, \addition{Lg5}, \addition{Lg6}). \addition{As such, we combine evaluation and deployment into one step, rather than separating the process into evaluation followed by deployment as other papers do~\cite{mlopsdefs, studer2021towards}.}
\addition{Lg3} described the multi-staged deployment process as follows:

\begin{displayquote}
    We have designated test clusters, [stage 1] clusters, [stage 2] clusters, then the global deployment [to all users].
    The idea here is you deploy increasingly along these clusters, so that you catch problems before they've met customers.
\end{displayquote}

Each organization had different names for its stages (e.g., test, dev, canary, staging, shadow, A/B)
and different numbers of stages in the deployment process (usually between one and four).
The stages helped invalidate models that might perform poorly in full production,
especially for brand-new or business-critical cases. Occasionally,
organizations had an offline ``sandbox'' stage preceding deployment to any fraction of
customers---for example, \addition{Md5} described a sandbox where they could stress-test their chatbot product:

\begin{displayquote}
    You can pretend to be a customer and say all sorts of offensive things, and see if the model will say cuss words back at you, or other sorts of things like that.
\end{displayquote}

Although the model retraining process was automated, we find that MLEs personally
reviewed validation metrics and manually supervised the promotion from one stage to the next.
They had oversight over every evaluation stage, taking great care to manage complexity
and change over time: \addition{specifically,} changes in data, product and business
requirements, users, and teams within organizations. We discuss
two human-centered practices: maintaining dynamic datasets and evaluating performance
in the context of the product or broader organizational value.

\subsubsection{\textbf{MLEs continuously update dynamic validation datasets}}\label{sec:summary-eval-dynamic}

Many engineers reported processes to analyze live failure modes and update
the validation datasets to prevent similar failures from happening again (\addition{Lg1}, \addition{Md1}, \addition{Lg2}, \addition{Lg3}, \addition{Sm3}, \addition{Md3}, \addition{Md5}, \addition{Sm6}, \addition{Md6}, \addition{Lg5}).
\addition{Lg1} described this process as a departure from what they had learned in school:

\begin{displayquote}
    You have this classic issue where most researchers are evaluat[ing] against fixed data sets [\dots but] most industry methods change their datasets.
\end{displayquote}

We found that these dynamic validation sets served two purposes:
(1) the obvious goal of making sure the validation set stays current with live data as much as possible,
given new knowledge about the problem and general shifts in the data distribution,
and (2) the more specific goal of addressing localized shifts within sub-populations,
such as low accuracy for minority groups. The challenge with (2) is that many sub-populations
are often overlooked, or they are discovered post-deployment~\cite{holstein_improving_2019}.
In response, \addition{Md3} discussed how they systematically bucketed their data points based on
the model's error metrics and created validation sets for each under-performing bucket:

\begin{displayquote}
    Some [of the metrics in my tool] are standard, like a confusion matrix,
    but it's not really effective because it doesn't drill things down [into specific subpopulations that users care about].
    Slices are user-defined, but sometimes it's a little bit more automated. [During offline evaluation, we]
    find the error bucket that [we] want to drill down, and then [we] either improve the model in very systematic ways or improve [our] data in very systematic ways.
\end{displayquote}

Rather than follow a proactive approach of constructing different failure modes in an offline validation phase
like \addition{Md3} did, \addition{Sm3} offered a reactive strategy of spawning a new dataset for each observed live failure:
\blockquote{Every [failed prediction] gets into the same queue, and 3 of us sit down once a week and go
    through the queue...then our [analysts] collect more [similar] data.}
This dataset update (or delta) was then merged into the validation dataset, and used for model validation in subsequent rounds.
While processes to dynamically update the validation datasets ranged from human-in-the-loop
to periodic synthetic data construction (\addition{Lg3}), we found that higher-stakes applications of ML (e.g., autonomous vehicles),
created dedicated teams to manage the dynamic evaluation process.
Often, this involved creating synthetic data representative of live failures (\addition{Lg1}, \addition{Lg3}, \addition{Md4}). For example, \addition{Lg1} said:

\begin{displayquote}
    What you need to be able to do in a mature MLOps pipeline is go very quickly from user recorded bug,
    to not only are you going to fix it, but you also have to be able to drive improvements to the stack by changing your data based on those bugs.
\end{displayquote}
Notwithstanding, participants found it challenging to collect the various kinds of failure modes and monitoring metrics for each mode. \addition{Lg6} added,
\blockquote{you have to look at so many different metrics. Even very experienced folks question this process like a dozen times.}

\subsubsection{\textbf{MLEs use product metrics for validation}}\label{sec:summary-eval-product}

While prior work discusses how prediction accuracy doesn't always correlate with real-world outcomes~\cite{autoai,deploychallenges},
it's unclear how to articulate clear and measurable ML goals. \citet{Patel2008InvestigatingSM}
discuss how practitioners trained in statistical techniques ``felt that they must often manage concerns outside the focus of traditional evaluation metrics.''
\citet{Srivastava2020AnEA} note that an increase in accuracy might not improve overall system
``compatibility.'' In our study, we found that successful ML deployments tied their performance to product metrics.
First, we found that \emph{prior to initial deployment}, mature ML teams defined a product metric in consultation
with other stakeholders, such as business analysts, product managers, or customers (\addition{Lg2}, \addition{Sm2}, \addition{Md5}, \addition{Sm6}, \addition{Md3}, \addition{Md6}, \addition{Lg5}, \addition{Lg6}).
Examples of product metrics include click-through rate and user churn rate. \addition{Md3} felt that a key reason many ML projects fail is that they don't measure metrics that will yield the organization value:

\begin{displayquote}
    Tying [model performance] to the business's KPIs (key performance indicators) is really important.
    But it's a process---you need to figure out what [the KPIs] are, and frankly I think that's how people should be doing AI.
    It [shouldn't be] like: hey, let's do these experiments and get cool numbers and show off these nice precision-recall
    curves to our bosses and call it a day. It should be like: hey, let's actually show the same business metrics that
    everyone else is held accountable to to our bosses at the end of the day.
\end{displayquote}

Since product-specific metrics are, by definition, different for different ML models, it was important for engineers to treat the choice of metrics as an explicit step in
their workflow and align with other stakeholders to make sure the right metrics were chosen.
After agreeing on a product metric, engineers only promoted experiment ideas to later deployment stages if there was an improvement in that metric.
\addition{Md6} said that every model change in production was validated by the product team:
\blockquote{if we can get a statistically significant greater percentage [of] people to subscribe to [the product], then [we can fully deploy].}
\citet{kim2016} also highlight the importance of including other stakeholders (or people in ``decision-making seats'') in the evaluation process.
At each stage of deployment, some organizations placed
additional emphasis on important customers during evaluation (\addition{Lg3}, \addition{Sm4}).
\addition{Lg3} mentioned that there were ``hard-coded'' rules for ``mission-critical'' customers:

\begin{displayquote}
    There's an [ML] system to allocate resources for [our product].
    We have hard-coded rules for mission critical customers. Like at the start of COVID, there were hospital [customers] that we had to save [resources] for.
\end{displayquote}

Finally, participants who came from research or academia noted that tying evaluation to the product metrics was a different experience.
\addition{Lg3} commented on their ``mindset shift'' after leaving academia:

\begin{displayquote}
    I think about where the business will benefit from what we're building.
    We're not just shipping fake wins, like we're really in the value business.
    You've got to see value from AI in your organization in order to feel like it was worth it to you,
    and I guess that's a mindset that we really ought to have [as a community].
\end{displayquote}

\subsection{Monitoring and Response}\label{sec:summary-monitoring}

\begin{displayquote}
    ``This data is supposed to have 50 states, there's only 40, what happened to the other 10?'' (\addition{Md6})
\end{displayquote}

We found that organizations centered their monitoring and response practices around engineers,
much like in the DevOps agile framework, which organizes software development around teams~\cite{cohen2004introduction}.
\addition{Prior work has stated that monitoring is critical to MLOps~\cite{mlopsdefs, mlopsoda, matsui2022mlops, studer2021towards}, and, broadly, that Agile practices can be useful in supervising production ML~\cite{amershi_software_2019}. We provide further insight by discussing two specific examples of Agile practices that our interviewees commonly adapted to the ML context.} First,
\addition{Lg3}, \addition{Lg4}, \addition{Md4}, \addition{Sm6}, \addition{Lg5}, and \addition{Lg6} described \emph{on-call processes} for supervising production ML models.
For each model, at any point in time, some ML engineer would be on call, or primarily responsible for it.
Any bug or incident observed (e.g., user complaint, pipeline failure) would receive a ticket, created by the on-call engineer.
On-call rotations typically lasted one or two weeks.
At the end of a shift, an engineer would
create an incident report---possibly one for each bug---detailing major issues that occurred and how they were fixed.
Additionally, \addition{Lg3}, \addition{Sm2}, \addition{Sm4}, and \addition{Md5} discussed having \emph{Service-Level Objectives (SLOs)}, or commitments to minimum standards of performance,
for pipelines in their organizations.
For example, a pipeline to classify images could have an SLO of 95\% accuracy. A benefit of using the
SLO framework for ML pipelines is a clear indication of whether a pipeline is performing well or not---if the SLO is not met, the pipeline is broken, by definition.

Our interviewees stressed the importance of logging data across all stages of the ML pipeline
(e.g., feature engineering, model training) to use for future debugging.
Monitoring ML pipelines and responding to bugs involved tracking live metrics (via queries or dashboards),
slicing and dicing sub-populations to investigate prediction quality, patching the model with non-ML heuristics for known failure modes,
and finding in-the-wild failures that could be added to future dynamic validation datasets.
While MLEs tried to automate monitoring and response as much as possible, we found that solutions were
lacking and required significant human-in-the-loop intervention. Next, we discuss data quality alerts, pipeline jungles, and diagnostics.

\subsubsection{\textbf{On-call MLEs track data quality alerts and investigate a fraction of them}}
\label{sec:summary-hacks-dataval}
In data science, data quality is of utmost importance~\cite{passi2018trust,kandel}.
Prior work has stressed the importance of monitoring data in production
ML pipelines~\cite{Sculley2015HiddenTD,sambasivan2021everyone,kim2016}, and the data management
literature has proposed numerous data quality metrics~\cite{dataval,polyzotis_data_2018,schelter}.
But what metrics do practitioners actually use, what data do practitioners monitor, and how do
they manually engage with these metrics? We found that engineers continuously monitored features for
and predictions from production models (\addition{Lg1}, \addition{Md1}, \addition{Lg3}, \addition{Sm3}, \addition{Md4}, \addition{Sm6}, \addition{Md6}, \addition{Lg5}, \addition{Lg6}):
\addition{Md1} discussed hard constraints for feature columns (e.g., bounds on values),
\addition{Lg3} talked about monitoring completeness (i.e., fraction of non-null values) for features,
\addition{Sm6} mentioned embedding their pipelines with "common sense checks," implemented as hard constraints on columns,
and \addition{Sm3} described schema checks---making sure each data item adheres to an expected set of columns and their types.
These checks were automated and executed as part of the larger pipeline (\Cref{sec:dag}).

While off-the-shelf data validation was definitely useful for the participants,
many of them expressed pain points with existing techniques and solutions. \addition{Lg3} discussed that it was hard to figure out
how to trigger alerts based on data quality:

\begin{displayquote}
    Monitoring is both metrics and then a predicate over those metrics that triggers alerts. That second piece doesn't exist---not
    because the infrastructure is hard, but because no one knows how to set those predicate values...for a lot of this stuff now,
    there's engineering headcount to support a team doing this stuff. This is people's jobs now; this constant, periodic evaluation of models.
\end{displayquote}
We also found that employee turnover makes data validation unsustainable (\addition{Sm2}, \addition{Md4}, \addition{Sm6}, \addition{Md6}, \addition{Lg5}).
If one engineer manually defined checks and bounds for each feature and then left the team,
another engineer would have trouble interpreting the predefined data validation criteria. To circumvent this problem,
some participants discussed using black-box data monitoring services but lamented that their
statistics weren't interpretable or actionable
(\addition{Sm2}, \addition{Md4}).

Another commonly discussed pain point was \emph{false-positive alerts}, or alerts triggered even when the ML performance is adequate.
Engineers often monitored and placed data quality alerts on each feature and prediction (\addition{Lg2}, \addition{Lg3}, \addition{Sm3}, \addition{Md3}, \addition{Md4}, \addition{Sm6}, \addition{Md6}, \addition{Lg5}, \addition{Lg6}).
If the number of metrics tracked grew too large, false-positive alerts could become a problem.
An excess of false-positive alerts led to fatigue and silencing of alerts, which could miss actual performance drops.
\addition{Sm3} said \blockquote{people [were] getting bombed with these alerts.} \addition{Lg5} shared a similar sentiment,
that there was \blockquote{nothing critical in most of the alerts.} The only time there was something critical
was \blockquote{way back when [they] had to actually wake up in the middle of the night to solve it...the only time [in years].}
When we asked what they did about the noncritical alerts and how they acted on the alerts, \addition{Lg5} said:

\begin{displayquote}
    You typically ignore most alerts...I guess on record I'd say 90\% of them aren't immediate. You just have to acknowledge them [internally], like just be aware that there is something happening.
\end{displayquote}

Seasoned MLEs thus preferred to view and filter alerts themselves, than to silence or lower the alert reporting rate.
In a sense, even false-positives can provide information about system health, if the MLE knows how to read the alerts and is accustomed to the system's reporting patterns.
When alert fatigue materialized, it was typically when engineers were on-call, or responsible for ML pipelines during a 7 or 14-day shift.
\addition{Lg6} recounted how on-call rotations were dreaded amongst their team, particularly for new team members, due to the high rate of false-positive alerts. They said:

\begin{displayquote}
    On-call ML engineers freak out in the first 2 rotations. They don't know where to look. So we have them act as a shadow, until they know the patterns.
\end{displayquote}

\addition{Lg6} also discussed an initiative at their company to reduce the alert fatigue, ironically with another model,
which would consider how many times an engineer historically acted on an alert of a given type before determining whether to surface a new alert of that type.

\subsubsection{\textbf{Over time, ML pipelines may turn into ``jungles'' of rules and models}}\label{sec:summary-monitoring-jungle}

\citet{Sculley2015HiddenTD} introduce the phrase ``pipeline jungles''
(i.e., different versions of data transformations and models glued together),
which was later adopted by participants in our study. While prior work demonstrates
their existence and maintenance challenges, we provide insight into why and how these pipelines become jungles in the first place.
Our interviewees noted that reacting to an ML-related bug in production usually took a long time, motivating
them to find strategies to quickly restore performance (\addition{Lg1}, \addition{Sm2}, \addition{Sm3}, \addition{Sm4}, \addition{Md4}, \addition{Md5}, \addition{Md6}, \addition{Lg6}).
These strategies primarily involved adding non-ML rules and filters to the pipeline.
When \addition{Sm3} observed, for an entity recognition task, that the model was misdetecting the
Egyptian president due to the many ways of writing his name, they thought it would be better to patch
the predictions for the individual case than to fix or retrain the model:

\begin{displayquote}
    Suppose we deploy [a new model] in the place of the existing model. We'd have to go through all the training data and then relabel it and [expletive], that's so much work.
\end{displayquote}

One way engineers reacted to ML bugs was by adding filters for models.
For the Egypt example, \addition{Sm3} added a simple string similarity rule to identify the president's name.
\addition{Md4} and \addition{Md5} each discussed how their models were augmented with a final,
rule-based layer to keep live predictions more stable.
For example, \addition{Md4} mentioned working on an anomaly detection model and adding a heuristics
layer on top to filter the set of anomalies that surface based on domain knowledge.
\addition{Md5} discussed one of their language models for a customer support chatbot:

\begin{displayquote}
    The model might not have enough confidence in the suggested reply, so we don't return [the recommendation]. Also, language models can say all sorts of things you don't necessarily want it to---another reason that we don't show some suggestions. For example, if somebody asks when the business is open, the model might try to quote a time when it thinks the business is open. [It might say] ``9 am,'' but the model doesn't know that. So if we detect time, then we filter that [reply]. We have a lot of filters.
\end{displayquote}

Constructing such filters was an iterative process---\addition{Md5} mentioned constantly stress-testing the model in a sandbox, as well as observing
suggested replies in early stages of deployment, to come up with filter ideas. Creating filters was a more effective strategy
than trying to retrain the model to say the right thing; the goal was to keep some version of a model working in production with little downtime.
As a result, filters would accumulate in the pipeline over time,
effectively creating a pipeline jungle. Even when models were improved,
\addition{Lg5} noted that it was too risky to remove the filters, since the filters were already in production,
and a removal might lead to cascading or unforeseen failures.

Several engineers also maintained fallback models for reverting to: either older or simpler versions (\addition{Lg2}, \addition{Lg3}, \addition{Md6}, \addition{Lg5}, \addition{Lg6}).
\addition{Lg5} mentioned that it was important to always keep some model up and running, even if they \blockquote{switched to a less economic model and had to just cut the losses.}
Similarly, when doing data science work, both \citet{passi2018trust} and \citet{autoai} echo the importance of having some solution to meet clients' needs, even if it is not the best solution.
Another simple solution engineers discussed was serving a separate model for each customer (\addition{Lg1}, \addition{Lg3}, \addition{Sm2}, \addition{Sm4}, \addition{Md3}, \addition{Md4}).
We found that engineers preferred a per-customer model because it minimized downtime:
if the service wasn't working for a particular customer, it could still work for other customers.
\citet{Patel2008InvestigatingSM} also noted that per-customer models could yield higher overall performance.

\subsubsection{\textbf{Bugs in production ML follow a heavy-tailed distribution}}\label{sec:summary-monitoring-bugs}
ML debugging is different from debugging during standard software engineering,
where one can write test cases to cover the space of potential bugs~\cite{deploychallenges, amershi_software_2019}.
\addition{Lg3}, \addition{Sm2}, \addition{Sm3}, \addition{Sm4}, \addition{Lg4}, \addition{Md4}, \addition{Md5}, \addition{Sm6}, \addition{Lg5}, and \addition{Lg6} mentioned having a \emph{central queue of production ML bugs}
that every engineer added tickets to and processed tickets from.
Often this queue was larger than what engineers could process in a timely manner, so they assigned tags to tickets to prioritize what to debug.

Interviewees discussed ad-hoc approaches to debugging production ML issues,
which could require them to spend a lot of time diagnosing any given bug (\addition{Lg3}, \addition{Lg2}, \addition{Sm3}, \addition{Sm4}, \addition{Lg5}).
One common issue was \emph{data leakage}---i.e., assuming during training that there is access
to data that does not exist at serving time---an error typically discovered after the model was
deployed and several incorrect live predictions were made
(\addition{Lg1}, \addition{Md1}, \addition{Md5}, \addition{Lg5}). Interviewees felt that anticipating all possible forms of data leakage during experimentation was tedious;
thus, sometimes leakage was retroactively checked during code review in an evaluation stage (\addition{Lg1}.
There were other types of bugs that were discussed by multiple participants, such as accidentally
flipping labels in classification models (\addition{Lg1}, \addition{Sm1}, \addition{Lg3}, \addition{Md3}) and forgetting to set random seeds
in distributed training when initializing workers in parallel (\addition{Lg1}, \addition{Lg4}, \addition{Sm5}).
However, the vast majority of bugs described to us in the interviews were seemingly bespoke and not shared among participants.
For example, \addition{Sm3} forgot to drop special characters (e.g., apostrophes) for their language models.
\addition{Lg3} found that the imputation value for missing features was once corrupted.
\addition{Lg5} mentioned that a feature of unstructured data type (e.g., JSON)
had half of the keys' values missing for a \blockquote{long time.}

When asked how they detect these one-off bugs, interviewees mentioned that their bugs showed similar symptoms of failure.
One symptom was a large discrepancy between offline validation accuracy and production accuracy immediately after deployment (\addition{Lg1}, \addition{Lg3}, \addition{Md4}, \addition{Lg5}).
However, if there were no ground-truth labels available immediately after deployment (as discussed in \Cref{sec:summary-dataprep-feedbackdelays}),
interviewees had to resort to other strategies. For example, some inspected the results of data quality checks (\Cref{sec:summary-hacks-dataval}).
\addition{Lg1} discussed their struggle to debug without ``ground-truth:'':

\begin{displayquote}
    Um, yeah, it's really hard. Basically there's no surefire strategy.
    The closest that I've seen is for people to integrate a very high degree of observability into every part of their pipeline.
    It starts with having really good raw data, observability, and visualization tools. The ability to query.
    I've noticed, you know, so much of this [ad-hoc bug exploration] is just---if you make the friction [to debug] lower,
    people will do it more. So as an organization, you need to make the friction very low for investigating what the data actually looks like, [such as] looking at specific examples.
\end{displayquote}

To diagnose bugs, interviewees typically sliced and diced data for different groups of customers
or data points (\addition{Md1}, \addition{Lg3}, \addition{Md3}, \addition{Md4}, \addition{Md6}, \addition{Lg6}). Slicing and dicing is known to be useful for identifying bias
in models~\cite{sambasivan2021everyone, holstein_improving_2019}, but we observed that our interviewees
used this technique beyond debugging bias and fairness; they sliced and diced to determine common failure modes and
data points similar to these failures. \addition{Md4} discussed annotating bugs and only drilling down into their
queue of bugs when they observed \blockquote{systematic mistakes for a large number of customers.}

Interviewees mentioned that after several iterations of chasing bespoke ML-related bugs in production,
they had developed a sense of paranoia while evaluating models offline---possibly as a coping mechanism (\addition{Lg1}, \addition{Md1}, \addition{Lg3}, \addition{Md5}, \addition{Md6}, \addition{Lg6}). \addition{Lg1} said:

\begin{displayquote}
    ML [bugs] don't get caught by tests or production systems and just silently cause errors. This is why [you] need to be paranoid when you're writing ML code, and then be paranoid when you're coding.
\end{displayquote}

\addition{Lg1} then recounted a bug that was \blockquote{impossible to discover} after a deployment to production:
the code for a change that added new data augmentation to the training procedure had two variables flipped,
and this bug was miraculously caught during code review even though the training accuracy was high.
\addition{Lg1} claimed that there was \blockquote{no mechanism by which [they] would have found this besides
    someone just curiously reading the code.} Since ML bugs don't cause systems to go down,
sometimes the only way to find them is to be cautious when inspecting code, data, and models.

\section{Discussion}\label{sec:discussion}

Our findings suggest that automated production ML pipelines are enabled by MLEs working through a continuous loop of
i) data preparation, ii) experimentation, iii) evaluation \& deployment, and iv) monitoring and response (\Cref{fig:tasksdiagram}).
Although engineers leverage different tools to help with technical aspects of their workflow,
such as experiment tracking and data validation~\cite{mlflow, dataval},
patterns began to emerge when we studied how MLE practices varied across company sizes and experience levels. We discuss these patterns as ``the three Vs of MLOps'' (\Cref{sec:discussion-3vs}),
and follow our discussion with implications for both production ML tooling (\Cref{sec:discussion-tools}), and opportunities for future work (\Cref{sec:discussion-limitations}).

\subsection{Velocity, Visibility, Versioning: Three Vs of MLOps}\label{sec:discussion-3vs}

Findings from our work and prior work suggest three broad themes of successful MLOps practices: Velocity, Visibility, and Versioning.
These themes have synergies and tensions across each stage of MLEs' workflow, as we discuss next.

\subsubsection{\textbf{Velocity}} Since ML is so experimental in nature, it's important to be able to prototype and iterate on ideas quickly
(e.g., go from a new idea to a trained model in a day). Interviewees attributed their productivity to development environments that
prioritized high experimentation velocity and debugging environments that allowed them to test hypotheses quickly.
Prior work has extensively documented the Agile tendencies of MLEs, describing how they iterate quickly
(i.e. with \emph{velocity}) to explore a large ML or data science search space~\cite{amershi_software_2019, demystifying, hohman2020understanding, Patel2008InvestigatingSM, zhang2020data}.
\citet{amershi_software_2019} describe how experimentation can be sped up when labels
are annotated faster (i.e., rapid data preparation). \citet{garcia2020flor} explore tooling to help MLEs correct logging oversights from too much velocity in experimentation,
and \citet{deploychallenges} mention the need to diagnose production bugs quickly to prevent future similar issues from occurring. First, our study re-enforces the
view the MLEs are agile workers who value fast results. P1 said that people who achieve the best outcomes from experimentation are people
with ``scary high experimentation velocity.'' Similarly, the multi-stage
deployment strategy can be viewed as an optimistic or high-velocity solution to deployment:
deploy first, and validate gradually across stages. Moreover, our study provides deeper insight into how practitioners
rapidly debug deployments---we identify and describe practices such as on-call rotations, human-interpretable filters on model behavior, data quality alerts, and model rollbacks.

At the same time, high velocity can lead to trouble if left unchecked.
\citet{sambasivan2021everyone} observed that, for high-stakes customers,
practitioners iterated too quickly, causing ML systems to fail---practitioners
``moved fast, hacked model performance (through hyperparameters rather than data quality),
and did not appear to be equipped to recognise upstream and downstream people issues.''
Our study exposed strategies that practitioners used to prevent themselves from iterating too quickly: for example,
in \Cref{sec:summary-eval-dynamic}, we described how some applications (e.g., autonomous vehicles)
require separate teams to manage evaluation, making sure that bad models don't get promoted from development to production.
Moreover, when measuring ML metrics outside of accuracy, e.g., fairness~\cite{holstein_improving_2019} or product metrics
(\Cref{sec:summary-eval-product}), it is challenging to make sure all metrics improve for each
change to the ML pipeline~\cite{deploychallenges}. Understanding which metrics to prioritize requires domain and business expertise~\cite{kim2016}, which could hinder velocity.

\subsubsection{\textbf{Visibility}}
In organizations, since many stakeholders and teams are impacted by ML-powered applications and services, it is important for
MLEs to have an end-to-end view of ML pipelines. P1 explicitly mentioned integrating ``very high degree of observability into every part of [the] pipeline'' (\Cref{sec:summary-monitoring-bugs}).
Prior work describes the importance of visibility: for example, telemetry data from ML pipelines (e.g., logs and traces)
can help engineers know if the pipeline is broken~\cite{kim2016}, model explainability methods can establish customers' trust
in ML predictions~\cite{deploychallenges,autoai,Klaise2020MonitoringAE}, and dashboards on ML pipeline health can help align
nontechnical stakeholders with engineers~\cite{kandel,krossorienting}.
In our view, the popularity of Jupyter notebooks among MLEs, including among the participants in our study,
can be explained by Jupyter's gains in velocity and visibility for ML experimentation, as it effectively combines REPL (Read-Eval-Print-Loop)-style
interaction and visualization capabilities despite its \emph{versioning} shortcomings.
Our findings corroborate these prior findings and provide further insight on how visibility is achieved in practice.
For example, engineers proactively monitor feedback delays (\Cref{sec:summary-dataprep-feedbackdelays}).
They also document live failures frequently to keep validation datasets current (\Cref{sec:summary-eval-dynamic}),
and they engage in on-call rotations to investigate data quality alerts (\Cref{sec:summary-monitoring}).

Visibility also helps with velocity. If engineers can quickly identify the source of a bug, they can fix it faster.
Or, if other stakeholders, such as product managers or business analysts, can understand how an experiment or multi-staged deployment is progressing,
they can better use their domain knowledge to assess models according to product metrics (see \Cref{sec:summary-eval-product}),
and intervene sooner if there's evidence of a problem.
One of the pain points we observed was that end-to-end experimentation---from the conception of an idea to improve ML performance to validation of the idea---took
too long. The uncertainty of project success stems from the unpredictable, real-world nature of experiments.

\subsubsection{\textbf{Versioning}} \citet{amershi_software_2019} mention that ``fast-paced model iteration'' requires careful versioning of data and code.
Other work identifies a need to also manage model versions~\cite{modeldb, mlflow}. Our work suggests that mananging \emph{all} artifacts---data,
code, models, data quality metrics, filters, rules---in tandem is extremely challenging but vital to the success of ML deployments.
Prior work explains how these artifacts can be queried during debugging~\cite{vizier,chavan2015towards,Sculley2015HiddenTD}, and our findings additionally show that versioning is particularly
useful when \emph{teams} of people work on ML pipelines. For instance, during monitoring, on-call engineers may receive a flood of false-positive alerts;
looking at old alerts might help them understand whether a specific type of alert actually requires action. In another example, during experimentation,
ML engineers often work on models and pipelines they didn't initially create.
Versioning increases visibility: engineers can inspect old versions of experiments to understand ideas that may or may not have worked.

Not only does versioning aid visibility, but it also enables workflows to maintain high velocity.
In \Cref{sec:summary-monitoring-jungle}, we explained how pipeline jungles are created by quickly responding to
ML bugs by constructing various filters and rules. If engineers had to fix the training dataset or model for every bug,
they would not be able to iterate quickly. Maintaining different versions for different types of
inputs (e.g., rules to auto-reject incomplete data or different models for different users) also enables high velocity.
However, there is also a tension between velocity and versioning: in \Cref{sec:summary-expt-iterative},
we discussed how parallelizing experiment ideas produces many versions, and ML engineers could not cognitively keep track of them.
In other words, having high velocity can mean drowning in a sea of versions.

\subsection{Opportunities for ML Tooling}\label{sec:discussion-tools}

Our main takeaway is that production ML tooling needs to aid \emph{humans} in their workflows,
not just automate technical practices (e.g., generating a feature or training a model).
Tools should help improve at least one of the three Vs, and there are opportunities for tools in each stage of the workflow.
We discuss each in turn.

\subsubsection{Data Preparation}

As mentioned in \Cref{sec:summary-dataprep}, separate teams of data engineers typically manage pipelines to ingest,
clean, and preprocess data on a schedule. While existing tools automate scheduling these activities,
there are unadressed ML needs around retraining and labeling.
Prior work and our interviews indicate that ML engineers retrain models on some arbitrary
cadence~\cite{deploychallenges, mlopsoda},
without understanding the effect of the cadence on the quality of predictions.
Models might be stale if they are retrained only monthly, or they might retrain using invalid or corrupt
data if they are retrained faster than the data is validated and cleaned (e.g., hourly).
Moreover, the optimal retraining cadence depends on the data, ML task, and amount of organizational resources,
such as compute, training time, and number of engineers on the team. New tools can help with these challenges
and determine the best retraining cadence for ML pipelines. With respect to labeling,
existing tools help with either labeling at scale~\cite{ratner2017snorkel} or labeling with high quality~\cite{krishnan_palm_2017},
but it is hard to achieve both. As a result, organizations have custom infrastructure and teams to resolve label mismatches,
apply domain knowledge, and reject incorrect labels. Labeling tools can leverage ensembling and add postprocessing filters to reject
and resolve incorrect and inconsistent labels. Moreover, they should track feedback delays and surface this information to users.

\subsubsection{Experimentation}

As discussed in \Cref{sec:summary-expt-iterative},
experiments are typically done in development environments and then promoted to production clusters during deployment.
The mismatch between the two (or more!) environments can cause bugs, creating an opportunity for new tools.
The development cluster should maximize iteration speed (velocity), while the production cluster should minimize end-user prediction latency~\cite{crankshaw2017clipper}.
Hardware and software can be different in each cluster, e.g., GPUs for training vs. CPUs for inference, and Python vs. C++, which makes this problem challenging.
New tools are prioritizing reproducibility---turning Jupyter notebooks into production scripts~\cite{nbslicer},
for instance---but should also standardize how engineers interact with experimentation workflows.
For example, while experiment tracking tools can literally keep track of thousands of experiments,
how can engineers sort through all these versions and actually understand what the best experiments are doing?
Our findings and prior work show that the experimental nature of ML and data science leads to undocumented tribal knowledge within
organizations~\cite{kim2017data, kandel}.
Documentation solutions for deployed models and datasets have been proposed~\cite{gebru2021datasheets, mitchell2019model},
but we see an opportunity for tools to help document \emph{experiments}---particularly, failed ones.
Forcing engineers to write down institutional knowledge about what ideas work or don't work slows them down,
and automated documentation assistance would be quite useful.

\subsubsection{Evaluation and Deployment}

Prior work has identified several opportunities in the evaluation and deployment space.
For example, there is a need to map ML metric gains to product or business gains~\cite{mlopsoda, Madaiochecklists, kim2016}.
Additionally, tools could help define and calculate subpopulation-specific performance metrics~\cite{holstein_improving_2019}. From our study, we have observed a need
for tooling around the multi-staged deployment process. With multiple stages, the turnaround time from experiment idea to having a full production deployment
(i.e., deployed to all users) can take several months. Invalidating ideas in earlier stages of deployment can increase overall, end-to-end velocity.
Our interviewees discussed how some feature ideas no longer make sense after a few months, given the nature of how user behaviors change,
which would cause an initially good idea to never fully and finally deploy to production.
Additionally, an organization's key product metrics---e.g., revenue or number of clicks---might change in the middle of a multi-stage deployment, killing the deployment.
This negatively impacts the engineers responsible for the deployment. We see this as an opportunity for new tools to streamline ML deployments in this multi-stage pattern,
to minimize wasted work and help practitioners predict the end-to-end gains for their ideas.

\subsubsection{Monitoring and Response}

Recent work in ML observability identifies a need for tools to give end-to-end visibility on ML pipeline behavior and debug ML
issues faster~\cite{Shankar2022TowardsOF, vizier}. Basic data quality statistics, such as missing data and type or schema checks,
fail to capture anomalies in the values of data~\cite{tfx, dataval,polyzotis_data_2018}.
Our interviewees complained that existing tools that attempt to flag anomalies in the values of data points produce too many false positives (\Cref{sec:summary-hacks-dataval}).
An excessive number of false-positive alerts, i.e., data points flagged as invalid even if they are valid, leads to two pain points:
(1) unnecessarily maintaining many model versions or simple heuristics for invalid data points, which can be hard to keep track of, and
(2) a lower overall accuracy or ML metric, as baseline models might not serve high-quality predictions for these invalid points.
Moreover, due to feedback delays, it may not be possible to track ML performance (e.g., accuracy) in real time.
What metrics can be reliably monitored in real time, and what criteria should trigger alerts to maximize precision and recall when identifying model performance drops?
How can these metrics and alerting criteria automatically tune themselves over time, as the underlying data changes? We envision this to be an opportunity for new data management tools.

Moreover, as discussed in \Cref{sec:summary-monitoring-jungle}, when engineers quickly respond to production bugs,
they create pipeline jungles. Such jungles typically consist of several versions of models, rules, and filters.
Most of the ML pipelines that our interviewees discussed were pipeline jungles.
This combination of modern model-driven ML and old-fashioned rule-based AI indicates a need for managing filters (and versions of filters)
in addition to managing learned models. The engineers we interviewed managed these artifacts themselves.

\subsection{Limitations and Future Work}\label{sec:discussion-limitations}

Since we wanted to find common themes in production ML workflows across different applications and organizations,
our interview study's scope was quite broad: we set out on a quest to discover \addition{shared patterns, rather than to predict or explain}.
We asked practitioners open-ended questions about their MLOps workflows and challenges,
but did not probe them about questions of fairness, risk, and data governance:
these questions could be studied in future interviews. \addition{Moreover, we did not focus on the \emph{differences} between practitioners' workflows based on their company sizes, educational backgrounds, or industries. While there are interview studies for specific applications of
ML~\cite{pumplun2021adoption,billington2022machine,folstad2018makes}, we see further opportunities to study the effect of organizational focus and maturity on the production ML workflow.}
There are also questions for which interview studies are a poor fit. Given our findings on the importance of collaborative and social dimensions of MLOps, we would like to explore these
ideas further through participant action research or contextual inquiry.

Moreover, our paper focuses on a \emph{human-centered} workflow surrounding production ML pipelines. Focusing on the \emph{automated} workflows
in ML pipelines---for example, continuous integration and continuous deployment (CI/CD)---could prove a fruitful research direction.
Finally, we only interviewed ML engineers, not other stakeholders, such as software engineers or product managers.
\citet{mlopsoda} present a diagram of technical components of the ML pipeline (e.g., feature engineering, model training)
and interactions between ML engineers and other stakeholders. Another interview study could observe these interactions and provide further insight into practitioners' workflows.

\section{Conclusion}

In this paper, we presented results from a semi-structured interview study of 18 ML engineers
spanning different organizations and applications to understand their workflow, best practices, and challenges.
Engineers reported several strategies to sustain and improve the performance of production ML pipelines,
and we identified four stages of their MLOps workflow:
i) data preparation, ii) experimentation, iii) evaluation and deployment, and iv) monitoring and response.
Throughout these stages, we found that successful MLOps practices center around having good velocity,
visibility, and versioning. Finally, we discussed opportunities for tool development and research.




\begin{acks}
We acknowledge support from grants DGE2243822, IIS-2129008, IIS-1940759, and IIS-1940757 awarded by the
National Science Foundation, an NDSEG Fellowship, funds from the Alfred P. Sloan Foundation, as well as EPIC lab sponsors: GResearch, Adobe, Microsoft, Google, and Sigma Computing.
\end{acks}

\bibliographystyle{ACM-Reference-Format}
\bibliography{sample-base}

\clearpage
\appendix

\section{Semi-Structured Interview Questions}
\label{app:interviewq}

In the beginning of each interview, we explained the purpose of the interview---to better understand processes within the organization for validating changes made to production ML models, ideally through stories of ML deployments.
We then kickstarted the information-gathering process with a question to build rapport with the interviewee,
such as \emph{tell us about a memorable previous ML model deployment}.
This question helped us isolate an ML pipeline or product to discuss. We then asked a series of open-ended questions:

\begin{enumerate}
  \item \textbf{Nature of ML task}
        \begin{itemize}
          \item What is the ML task you are trying to solve?
          \item Is it a classification or regression task?
          \item Are the class representations balanced?
        \end{itemize}
  \item \textbf{Modeling and experimentation ideas}
        \begin{itemize}
          \item How do you come up with experiment ideas?
          \item  What models do you use?
          \item How do you know if an experiment idea is good?
          \item  What fraction of your experiment ideas are good?
        \end{itemize}
  \item \textbf{Transition from development to production}
        \begin{itemize}
          \item What processes do you follow for promoting a model from the development phase to production?
          \item How many pull requests do you make or review?
          \item What do you look for in code reviews?
          \item What automated tests run at this time?
        \end{itemize}
  \item \textbf{Validation datasets}
        \begin{itemize}
          \item How did you come up with the dataset to evaluate the model on?
          \item Do the validation datasets ever change?
          \item Does every engineer working on this ML task use the same validation datasets?
        \end{itemize}
  \item \textbf{Monitoring}
        \begin{itemize}
          \item Do you track the performance of your model?
          \item If so, when and how do you refresh the metrics?
          \item What information do you log?
          \item Do you record provenance?
          \item How do you learn of an ML-related bug?
        \end{itemize}

  \item \textbf{Response}
        \begin{itemize}
          \item What historical records (e.g., training code, training set) do you inspect in the debugging process?
          \item What organizational processes do you have for responding to ML-related bugs?
          \item Do you make tickets (e.g., Jira) for these bugs?
          \item How do you react to these bugs?
          \item When do you decide to retrain the model?
        \end{itemize}
\end{enumerate}

\addition{\section{Tools Referenced in Interviews}}
\label{app:tools}

\addition{\Cref{tab:tools} lists several of the tools that were commonly referenced by the interviewees.}

\begin{table*}[t!]
    \centering
    \resizebox{\linewidth}{!}{%
    \textcolor{revisionBlue}{\noindent\begin{tabular}{|r|p{0.18\linewidth}|p{0.18\linewidth}|p{0.18\linewidth}|p{0.18\linewidth}|}
    \toprule
       & \multicolumn{1}{|c|}{\textbf{Data Collection}} & \multicolumn{1}{|c|}{\textbf{Experimentation}}  &  \multicolumn{1}{|c|}{\textbf{Evaluation and Deployment}} & \multicolumn{1}{|c|}{\textbf{Monitoring and Response}} \\
    \midrule
    \textbf{Metadata} & {Data catalogs, Amundsen, AWS Glue, Hive metastores} & \multicolumn{2}{p{0.36\linewidth}|}{{Weights \& Biases, MLFlow, train/test set parameter configs, A/B test tracking tools}} & {Dashboards, SQL, metric functions and window sizes} \\
     \midrule
    \textbf{Unit} & {Data cleaning tools} & {Tensorflow, MLlib, PyTorch, Scikit-learn, XGBoost} & {OctoML, TVM, joblib, pickle} & {Scikit-learn metric functions, Great Expectations, Deequ} \\
    \cmidrule(lr){2-5} 
    & \multicolumn{4}{c|}{{Python, Pandas, Spark, SQL, C++, ONNX}} \\
     \midrule
      \textbf{Pipeline} & {In-house or outsourced annotators} & {AutoML} & {Github Actions, Travis CI, Prediction serving tools, Kafka, Flink} & {Prometheus, AWS CloudWatch} \\
    \cmidrule(lr){2-5}
    & \multicolumn{4}{c|}{{Airflow, Kubeflow, Argo, Tensorflow Extended (TFX), Vertex AI, DBT}} \\
     \midrule
    \textbf{Infrastructure} & {Annotation schema, cleaning criteria configs} & {Jupyter notebook setups, GPUs} & {Edge devices, CPUs} & {Logging and observability services (e.g., DataDog)} \\
    \cmidrule(lr){2-5}
    & \multicolumn{4}{c|}{{Cloud (e.g., AWS, GCP), compute clusters, storage (e.g., AWS S3, Snowflake), Docker, Kubernetes}} \\
     \bottomrule
    \end{tabular}%
    }}
    \caption{\textcolor{revisionBlue}{Common tools referenced in interview transcripts, segmented by stage in the MLOps workflow and layer in the stack. The metadata layer is concerned with artifacts for component runs, like results of a training script. The unit layer represents individual pieces or components of a pipeline, such as feature engineering or model training. The pipeline layer connects components through orchestration frameworks, and the lowest layer is the infrastructure (e.g., compute).}}
    \label{tab:tools}
\end{table*}


\end{document}